\documentclass{aa}  
\usepackage[colorlinks, citecolor=blue]{hyperref}
\usepackage{graphicx}
\usepackage{txfonts}
\usepackage{lscape}
\usepackage{siunitx}
\def\lsol{L$_{\odot}$}
\def\Msun{M$_{\odot}$}
\def\msun{M$_{\odot}$}

\def\rsun{R$_{\odot}$}

\begin{document}

\title{Probing the low-mass end of the  companion mass function for O-type stars}

\author{M. Reggiani\inst{1}, A. Rainot\inst{1,2}, H. Sana\inst{1}, L. A. Almeida\inst{3,4}, S. Caballero-Nieves\inst{5}, K. Kratter\inst{6}, S. Lacour\inst{7}, J.-B. Le Bouquin\inst{8}, H. Zinnecker\inst{9}}

   \institute{Institute of astrophysics, KU Leuven, Celestijnlaan 200D, 3001 Leuven, Belgium\\
              \email{maddalena.reggiani@kuleuven.be}
        \and
              ScanWorld SA, Rue des Chasseurs Ardennais 6, Li\`ege Science Park, 4031 Angleur, Belgium
        \and
              Escola de Ci\^encias e Tecnologia, Universidade Federal do Rio Grande do Norte, Natal, RN 59072-970, Brazil
        \and
               Programa de P\'os-gradua\c{c}\~ao em F\'isica, Universidade do Estado do Rio Grande do Norte, Mossor\'o, RN, 59610-210, Brazil
        \and
               Department of Aerospace Physics \& Space Sciences, Florida Institute of Technology 150 West University Blvd, Melbourne, FL 32901, USA
        \and
               Department of Astronomy, University of Arizona, Tucson, AZ 85721, USA
        \and
               LESIA, (UMR 8109), Observatoire de Paris, PSL, CNRS, UPMC, Universit\'e Paris-Diderot, 5 place Jules Janssen, 92195 Meudon, France
        \and
                Institut de Plan\'etologie et d’Astrophysique de Grenoble Universit\'e, Grenoble, 38058 Grenoble, France
        \and
               Universidad Aut\'onoma de Chile, Avda Pedro de Valdivia 425, Providencia, Santiago de Chile, Chile
             }
   \date{}

   \authorrunning{M.~Reggiani et al.}


 
  \abstract
   {Past observations of O-type stars in the Galaxy have shown that almost all massive stars are part of a binary or higher-order multiple system. Given the wide range of separations at which these companions are found, several  observational techniques have been adopted to characterize them. Despite the recent advancements in interferometric and adaptive optics observations, contrasts greater than 4 in the $H$ band have never been reached between 100 and 1000 mas.}
   {Using new adaptive optics (AO) assisted coronagraphic observations, we aim to study the multiplicity properties of a sample of 18 dwarf (or sub-giant) O stars in the galactic field and in OB associations to probe the existence of stellar companions in the angular separation range from 0\farcs15 to 6\arcsec~ down to very low mass ratios.}
   {We  used VLT/SPHERE to observe simultaneously with the IRDIS and IFS sub-systems 18 O-type stars within 6 kpc and ages between 1-5 Myrs. The IFS $YJH$ band observations have allowed us to probe the presence of sub-solar companions in a 1.7\arcsec$\times$1.7\arcsec~field-of-view  down to magnitude limits of $\Delta H$=10 at 0\farcs4. In the wider 12\arcsec$\times$12\arcsec~ IRDIS field-of-view, we reached contrasts of $\Delta K$=12 at 1\arcsec, enabling us to look for even fainter companions at larger angular separations and to probe the source density of the surrounding portion of the sky.}
   {This paper presents five newly discovered intermediate ($<$1\arcsec) separation companions, three of which are smaller than 0.2 \Msun. If confirmed by future analyses of proper motions, these new detections represent the lowest-mass companions ever found around O-type stars. Additionally, 29  other sources  are found in the IRDIS field-of-view with spurious association probabilities smaller than 5\%. Assuming that all sources detected within 1\arcsec~are physically bound companions, the observed (uncorrected for bias) fraction of companions for O-type stars between 150 and 900 mas is $0.39\pm0.15$, whereas it increases to $1.6\pm0.3$ in the separation range from 0\farcs9 to 6\arcsec.}
   {These findings clearly support the notion that massive stars form almost exclusively in multiple systems, serving as proof of concept that supports the application of larger AO-assisted coronagraphic surveys as a crucial step in placing constraints on the multiplicity properties of massive star companions in regions of the parameter space that have previously gone unexplored. These results also demonstrate that the companion mass function is populated down to the lowest stellar masses.}


   \maketitle
%

\section{Introduction}
The formation of massive stars remains an open question in astronomy today. For low-mass stars, the core accretion paradigm -- starting with the collapse of pre-stellar cores into a single or binary protostar with the subsequent accretion of matter through a Keplerian disk \citep[e.g., ][]{Shu1987,Inutsuka2012} -- is expected to produce stars over timescales of 10-50 Myr. The problems in scaling up the low-mass star formation models to the high-mass regime  indeed come from the short timescales involved \citep[e.g.,][]{Zinnecker2007,Brott2011}. Therefore, for massive stars, other mechanisms such as competitive accretion \citep{Bonnell2001}  and protostellar mergers \citep{Bonnell1998,Bally2005,Moeckel2011} have been proposed. Except for collisional models, most high-mass star formation theories are in agreement with regard to the presence of dense and massive accretion disks in sustaining accretion in the presence of radiation pressure. These disks are presumably unstable to fragmentation and  this can lead to the formation of stellar companions \citep{Kratter2011}.

\begin{table*}[t]
 \caption[]{\label{Table:O star properties} List of O-type stars used in this work.}
\begin{tabular}{lccccc}
 \hline \hline
  Object ID &
  SpT &
 d (kpc) & Age & Association &
  References \\
 \\ \hline
HD~\num{64568}   &  O3~V((f*))z &  4.25$^{+0.21}_{-0.19}$ &  & NGC 2467, Pup OB1 & 1,2  \\
HD~\num{93128}   & O3.5~V((fc))z & 2.363$^{+0.061}_{-0.058}$ & 0.3-0.5 Myrs & Trumpler 14, Car OB1 & 1,2,6 \\
HD~\num{155913}   &  O4.5~Vn((f)) &   1.27$^{+0.14}_{-0.12}$ &  & RCW 114 &  2,3\\
HDE~\num{319699}   & O5~V((fc)) & 1.67$^{+0.19}_{-0.16}$ & 1.9 Myr & NGC 6334,RCW 127 & 2,3,9 \\
HD~\num{124314}   & O6~IV(n)((f)) & 1.75$^{+0.20}_{-0.16}$  &   & -- &  2, 3 \\
HD~\num{150135}   &  O6.5~V((f))z &  1.148$^{+0.016}_{-0.015}$  & 1–5 Myrs & NGC 6193, Ara OB1a & 1,13,8 \\
HDE~\num{326775}   & O6.5~V(n)((f))z &  1.95$^{+0.39}_{-0.28}$  &  & RCW 113 & 3,14,11 \\
V3903~Sgr   &  O7~V(n)z + B0:~V & 1.08$^{+0.10}_{-0.08}$ &  & Sgr OB1 & 2,3  \\
HD~\num{152623}   &  O7~V(n)((f)) & 1.16 & 2–4 Myr & Trumpler 24, Sco OB1 & 12,14 \\
BD~$-$13\degr~4929   & O8~V + B0.5:~V + B0.5:~V &  1.697$^{+0.031}_{-0.030}$ & 1-2 Myrs & NGC 6611 &  1,5,4 \\
HD~\num{101191}   & O8~V & 2.82$^{+0.67}_{-0.46}$ &  & IC 2944, Cru OB1 & 2,3,7 \\
HDE~\num{323016}   & O8.5~V &  2.13$^{+0.34}_{-0.26}$ &  & -- & 3,10 \\
HD~\num{149452}   & O9~IVn &  1.34$^{+0.14}_{-0.11}$ &  & RCW 108, Ara OB1ab & 2,3  \\
HD~\num{76341}   &  O9.2~IV &   1.33$^{+0.19}_{-0.15}$  & & RCW, VelOB1  &   2,3 \\
BD~$-$13\degr~4928   & O9.5 V &  1.697$^{+0.031}_{-0.030}$  & 1-2 Myrs & NGC 6611 & 1, 4 \\
CPD~$-$41\degr~7721   &  O9.7~V:(n) & 1.551$^{+0.025}_{-0.024}$  & 2–4 Myr & NGC 6231, Sco OB1 & 1, 14 \\
HD~\num{123056}   &  O9.5~IV(n) & 1.53$^{+0.18}_{-0.14}$  &  & -- &  2,3 \\
HD~\num{152200}   &  O9.7~IV(n) &  1.551$^{+0.025}_{-0.024}$  & 2–4 Myr & NGC 6231,  Sco OB1 & 2,3 \\
\hline
\end{tabular}
\tablebib{(1)~\citet{MaizApellaniz2021};
(2) \citet{Sota2014}; (3) \citet{PantaleoniGonzalez2021}; (4) \citet{Sana2009};
(5) \citet{Martayan2008}; 
(6) \citet{Sana2010}; (7) \citet{Sana2011}; (8) \citet{Baume2011}; (9) \citet{Russeil2017};
(10) \citet{Neckel1981}; (11) \citet{Arias2016}; (12) \citet{Shull2019}; (13) \citet{MaizApellaniz2020}; (14) \citet{MaizApellaniz2016}.
}
\end{table*}

A clear observational evidence is that the multiplicity frequency is significantly higher among massive stars \citep{Duchene2013}. Several spectroscopic surveys of OB stars both in our Galaxy, and in the LMC \citep[see ee.g.,][]{Chini2012,Kobulnicky2012,Sana2012,Sota2014,Sana2013} have shown that the binary (or multiple) frequency may be $>$70\% for
binaries with physical orbital separation smaller than 1 AU.
The occurrence of longer period binaries has been explored through speckle interferometry by \citet{Mason1998,Mason2009}, adaptive optics
(AO) by \citet{Turner2008} and \citet{Close2012}, and lucky imaging by \citet{MaizApellaniz2010} and \citet{Peter2012}. These studies also demonstrate the high incidence of binaries and multiples among longer-period systems. In order to fill the observational gap between classical imaging and spectroscopic surveys, the Southern MAssive Stars at High angular resolution survey \citep[SMaSH+, ][]{Sana2014} has combined optical interferometry (VLTI/PIONIER) and aperture masking (NACO/SAM). The SMaSH+ survey is sensitive to mostly bright companions ($\Delta H < 4$) between 0\farcs001-0\farcs2 for a large sample of O-type stars. The occurrence of fainter ($\Delta H < 8$) companions at larger separations (up to 8\arcsec) was also probed in entire NACO field-of-view (FoV).
The main conclusion of this work was that nearly all massive stars have at least one
companion in the separation range covered by the observations and that over 60\%\ are part of  a higher order multiple.

It is thus critical to probe the frequency of even  fainter ($\Delta H < 4$) companions at these separations to determine the shape of the period and mass ratio distribution and to estimate the total binary frequency. The properties of the binary population may indeed serve as a useful diagnostics tool to discriminate between different formation models, particularly at separations that approximately correspond to the  size of the accretion disk, where we expect to find the low-mass companions formed from the remnants of the fragmented disk.

In this regard, the first paper of the Carina High-contrast Imaging Project of massive Stars \citep[CHIPS,][]{Rainot2020} represents a proof of concept that the extreme AO, implemented at the VLT through the Spectro-Polarimetric High-contrast Exoplanet REsearch instrument \citep[SPHERE,][]{Beuzit2019}, provides the necessary spatial resolution and dynamics to look for the faintest companions to nearby massive stars. 

In this paper, we analyze the multiplicity properties of a small sample of 18 dwarf O-type stars from the Galactic field, loose associations, or denser clusters to have a first statistics on the occurrence of faint ($\Delta K < 12$) companions in the $0\farcs15$ to $6\arcsec$~angular separation regime. 

First, we describe the sample we considered (Sect.~\ref{Sect:sample}). The setup for the observations and the data reduction are presented in Sect.~\ref{Sec:obs and data}. The image post processing is described in Sect.~\ref{Sec:image processing}. Then, in Sect.~\ref{Sec:results} we show our results and we discuss them in Sect.~\ref{Sec:discussion}.
Finally, we offer our conclusions in Sect~\ref{Sec:conclusions}.

\section{Sample}\label{Sect:sample}
Our sample consists of 18 dwarf O stars with spectral types ranging from O3~V to O9.7~V. Dwarf O stars only represented 20\% of the SMaSH+ sample due to the magnitude-limited (H <7.5) quality of the survey. These represent, however, the most natural targets in the search for observational constraints on the outcome of the massive star formation processes because they are less evolved than giants and supergiants and because they are intrinsically less luminous, allowing us to probe the low-mass end of the companion mass function. Thus, in the framework of this small-scale project, we exclusively targeted a sample of dwarf O stars covering a range of environments (clusters, diffuse OB associations, field) and masses (from 15 to 60 \msun). The properties of the objects are described in Table~\ref{Table:O star properties}. The ages of the targets are between and 1  to 5 Myrs and the distances from $\sim$ 1 to 4 kpc.

\section{Observations and data reduction}\label{Sec:obs and data}
The 18 O type stars were observed as part of a SPHERE Science Verification program and a standard service mode program in 2015. The high-contrast imaging instrument SPHERE is  mounted at the Naysmith platform of Unit 3 telescope (UT3) at ESO’s VLT, and consists of an extreme adaptive optics system, coronagraphic masks, and three different sub-systems. The observations were carried out in the IRDIS and IFS extended mode (IRDIFS\_EXT) by simultaneously using  the Integral Field Spectrograph (IFS) and the Infra-Red Dual-beam Imaging and Spectroscopy (IRDIS) sub-systems \citep{Galicher2018}. 

The IFS images have a pixel scale of 7.4 mas and with a total size of 290 $\times$ 290 pixels cover a 1\farcs73\ $\times$ 1\farcs73 \ FoV on the sky. 
IRDIS instead has a FoV of 12\arcsec$\times$12\arcsec~with a pixel scale of 12.25 mas (i.e., 1024 $\times$ 1024 pixels in total).
The IRDIFS\_EXT mode allows us to combine the $YJH$ band observations with IFS to dual $K$-band observations with IRDIS. Due to the size of the FoV and its spectroscopic capabilities, IFS enables the detection and characterizion of companions at close separations, whereas the larger FoV of IRDIS provides statistics for companions at larger separations and for the local field density of objects. 

All observations were carried out in pupil-tracking mode to allow for image post-processing through angular differential imaging \citep[ADI,][]{marois2006} techniques. However, most of the objects were not observed during meridian passage and only a limited parallactic angle variation was achieved.

For both IRDIS and IFS, the observing sequence was composed of three types of observations.  Our science frames ({\sc object, O}) were obtained by blocking the light coming from the bright central stars with SPHERE’s apodized Lyot coronagraphs. We also obtained {\sc center (C)} frames, which were acquired by applying a sinusoidal pattern to the deformable mirror to infer the position of the star behind the coronagraph. Finally, for spectro-photometric calibration, we took uncoronagraphic {\sc flux (F)} images of the stellar point spread function (PSF) by offsetting the central star from the coronagraphic mask and we used a neutral density filter (ND2.0) to avoid any saturation of the detector. 
The same F-C-O sequence was repeated three times for each target. For HD~\num{123056} only IFS observations were obtained.

The choice of detector integration times (DITs) and number of DITs (NDIT) for our {\sc object}  and {\sc flux} exposures  for each target are presented in Table \ref{table:Obs IFS} and in Table \ref{table:Obs IRDIS} for IFS and IRDIS, respectively.  In total, for every star, we obtained four-dimensional (4D) IFS and IRDIS data cubes. The IFS cubes are  composed of 290$\times$290 pixel images for each of the 39 wavelengths channels (from 0.9 to 1.6 $\mu$m) and each sky rotation. The IRDIS data cubes contain 1024$\times$1024 pixel images for each one of the two wavelengths channels (K1 and K2) and each sky rotation. The summary of the observing conditions and total parallactic angle variation for each object and is also presented in Tables \ref{table:Obs IFS} and \ref{table:Obs IRDIS}.

 \begin{table*}[t]
 \caption[]{\label{table:Obs IFS} Observing setup and atmospheric conditions for {\sc flux} (F) and {\sc object} (O) IFS observations.  Date corresponds to  the start of the exposures. Values of the seeing and the coherence time $\tau_{0}$ are the average values taken during the observations. }
\begin{tabular}{lccccccccc}
 \hline \hline
  Object ID & date & NDIT (O) & DIT (O)  & NDIT (F) & DIT (F) & Airmass  & PA variation  & Seeing & $\tau_{0}$ \\
    &   &   & [s] &  & [s] &   & (\degr) &  &  [s] \\
 \hline 
HD~\num{64568}  & 2014-12-08 & 8 & 16  & 64 & 2 & 1.00 & 5.6 & 1.45 & 0.002\\
HD~\num{93128}   & 2015-04-12 & 10 & 16 & 8 & 16 & 1.22 & 14.6 & 0.81 & 0.003 \\
HD~\num{155913}   & 2015-07-17 & 6 & 16 & 16 & 4 & 1.15 & 2.8 & 1.0 & 0.002\\
HDE~\num{319699}   & 2015-07-12  & 8 & 16 & 16 & 8 & 1.03 & 14.6 & 0.98 & 0.003 \\
HD~\num{124314}   & 2015-06-10 & 4 & 16 & 32 & 2 & 1.25 & 4.7 & 1.51 & 0.002 \\
HD~\num{150135}   & 2015-07-26 & 2 & 32 & 4 & 16 & 1.10 & 7.3  & 1.18 & 0.001\\
HDE~\num{326775}   & 2015-07-12 & 10 & 16 & 8 & 16 & 1.05 & 14.5 & 1.02 & 0.001\\
V3903~Sgr   & 2015-06-20 & 6 & 16 & 16 & 4 & 1.45 & 0.8 & 0.71 & 0.002 \\
HD~\num{152623}   & 2015-07-12 & 4 & 16 & 32 & 2 & 1.04 & 13.7 & 1.21 & 0.001 \\
BD~$-$13\degr~4929   & 2015-06-20 & 10 & 16 & 16 & 16 & 1.31 & 1.1 & 0.86 & 0.002 \\
HD~\num{101191}   & 2015-06-10 & 5 & 32 & 16 & 16 & 1.28 & 4.94  & 1.72 & 0.002 \\
HDE~\num{323016}   & 2015-07-17 & 8 & 16 & 16 & 8 & 1.19 & 4.1 & 0.81 & 0.002\\
HD~\num{149452}   & 2015-07-21 & 8 & 16  & 16 & 8 & 1.24 & 4.8 & 1.6  & 0.001\\
HD~\num{76341}   & 2014-12-09  & 12 & 16 & 32 & 4 & 1.09 & 10.2 & 0.7  & 0.003 \\
BD~$-$13\degr~4928   &2015-06-21  & 18 & 16 & 32 & 2 & 1.17 & 2.6 & 1.09 & 0.001 \\
CPD~$-$41\degr~7721 & 2015-08-23  & 10 & 16 & 16 & 16 & 1.14  & 8.6 & 0.98 & 0.007\\
HD~\num{123056}   & 2015-07-31 &  10 & 16 & 8 & 16 & 1.26 & 7.3  & 1.37  & 0.002 \\
HD~\num{152200}   & 2015-08-19 & 10 & 16 & 16 & 16 & 1.29 & 14.1 & 1.41 & 0.002 \\

\hline
\end{tabular}
\end{table*}
 
 \begin{table*}[t]
 \caption[]{\label{table:Obs IRDIS} Observing setup and atmospheric conditions for {\sc flux} (F) and {\sc object} (O) IRDIS observations.   Date corresponds to  the start of the exposures. Values of the seeing and the coherence time $\tau_{0}$ are the average values taken during the observations. }
\begin{tabular}{lccccccccc}
 \hline \hline
  Object ID & date & NDIT (O) & DIT (O)  & NDIT (F) & DIT (F) & Airmass  & PA variation  & Seeing & $\tau_{0}$ \\
    &   &   & [s] &  & [s] &   & (\degr) &  &  [s] \\
\hline
HD~\num{64568}  & 2014-12-08 & 1 & 16  & 4 & 4 & 1.00 & 8.0 & 0.84 & 0.003\\
HD~\num{93128}   & 2015-04-12 & 2 & 16 & 8 & 16 & 1.23 & 13.3 & 0.7 & 0.004 \\
HD~\num{155913}   & 2015-07-17 & 2 & 8 & 16 & 4 & 1.15 & 2.7 & 1.0 & 0.002\\
HDE~\num{319699}   & 2015-07-12  & 1 & 16 & 16 & 4 & 1.03 & 15.7 & 0.75 & 0.002 \\
HD~\num{124314}   & 2015-06-10 & 2 & 4 & 16 & 2 & 1.25 & 4.7 & 1.34 & 0.002 \\
HD~\num{150135}   & 2015-07-26 & 2 & 4 & 8 & 2 & 1.21 & 4.2  & 0.82 & 0.003\\
HDE~\num{326775}   & 2015-07-12 & 2 & 16 & 8 & 4 & 1.05 & 14.2 & 1.01 & 0.001\\
V3903~Sgr   & 2015-06-20 & 2 & 8 & 16 & 4 & 1.50 & 0.8 & 0.95 & 0.002 \\
HD~\num{152623}   & 2015-07-12 & 2 & 4 & 8 & 2 & 1.04 & 13.6 & 1.21 & 0.001 \\
BD~$-$13\degr~4929   & 2015-06-20 & 2 & 16 & 16 & 16 & 1.37 & 1.0 & 0.82 & 0.002 \\
HD~\num{101191}   & 2015-06-10 & 1 & 32 & 32 & 8 & 1.29 & 4.75  & 1.61 & 0.002 \\
HDE~\num{323016}   & 2015-07-17 & 1 & 16 & 16 & 8 & 1.21 & 3.9 & 1.00 & 0.002\\
HD~\num{149452}   & 2015-07-21 & 1 & 16  & 16 & 2 & 1.28 & 4.5 & 1.7  & 0.001\\
HD~\num{76341}   & 2014-12-09  & 4 & 8 & 16 & 2 & 1.08 & 9.8 & 0.67  & 0.004 \\
BD~$-$13\degr~4928   &2015-06-21  & 4 & 16 & 32 & 2 & 1.21 & 2.6 & 0.92 & 0.002 \\
CPD~$-$41\degr~7721 & 2015-08-23  & 2 & 16 & 16 & 16 & 1.14  & 8.4  & 0.89 & 0.008\\
HD~\num{123056}   & -- &  -- & -- & -- & -- & --  & --  & --  & -- \\
HD~\num{152200}   & 2015-08-19 & 2 & 16 & 16 & 16 & 1.30 & 13.9 & 1.47 & 0.002 \\
\hline
\end{tabular}
\end{table*}

The data reduction of IRDIS and IFS images was carried out by the SPHERE Data centre  \citep[DC]{Delorme2017} at the Institut de Planetologie et d'Astrophysique de Grenoble (IPAG)\footnote{\url{http://ipag.osug.fr/?lang=en}}. The SPHERE-DC applies a standard data reduction to the science and PSF frames by removing bad pixels, dark and flat frames and estimating the bias in each exposure. They calibrated the astrometry using the on-sky calibrations from \citet{Maire2016}, with a true north correction value of $1.75\pm0.08^\circ$ and a plate scale of $7.46\pm0.02$ mas/pixel and $12.255\pm0.009$ mas/pixel for IFS and IRDIS,  respectively. 

As three uncoronagraphic PSF observations were taken during our observing sequence at each of the wavelength channels (2 for IRDIS and 39 for IFS), we computed the median of the three PSF frames for each wavelength to increase the signal-to-noise ratio (S/N). We measured the total flux of the central object on the median-combined images at each wavelength and its uncertainty by computing the standard deviation of the flux measured in the three PSF frames.

\section{PCA image processing}\label{Sec:image processing}
\begin{figure*}
   \centering
   \includegraphics[width=19cm]{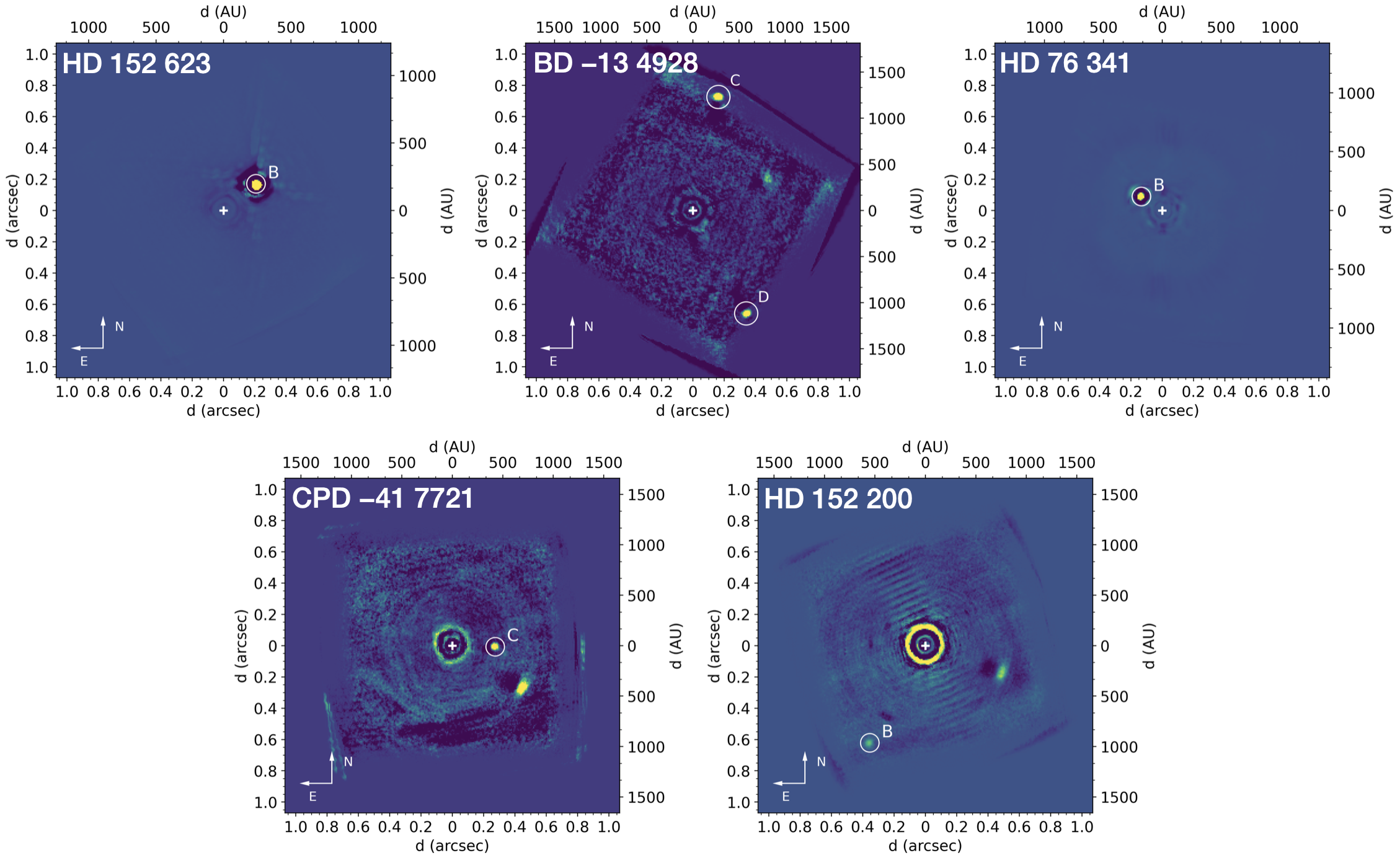}
      \caption{Final PCA/SDI IFS images for the targets with a detected source. The position of the central star is indicated by a white cross.}
         \label{Fig: final IFS images}
   \end{figure*}

To post-process the reduced data cubes we used the python open-source Vortex Imaging Processing package\footnote{\url{https://github.com/vortex-exoplanet/VIP}} \citep[VIP,][]{Gomez2016}, which was developed to analyze high-contrast imaging datasets for exoplanet detection. Based on the type of data and user choice, it also performs angular, reference, and spectral differential imaging (ADI, RDI, SDI, respectively), or simultaneous ADI+SDI, all based on a principal component analysis \citep[PCA, ][]{AmaraQuanz2012,Soummer2012} approach. 
In our study, for each object, we applied PCA/ADI separately to the two K1 and K2 IRDIS cubes and PCA/SDI on all IFS channels simultaneously to get the final post processed images. 
The resulting reduced PCA/ADI $K$1-band IRDIS frames for all our targets  are shown in Figure~\ref{Fig: final IRDIS images}. Only the final PCA/SDI IFS images that present possible companion detections are presented in Figure~\ref{Fig: final IFS images}.

   \begin{figure*}
   \centering
   \includegraphics[angle=0, width=18cm]{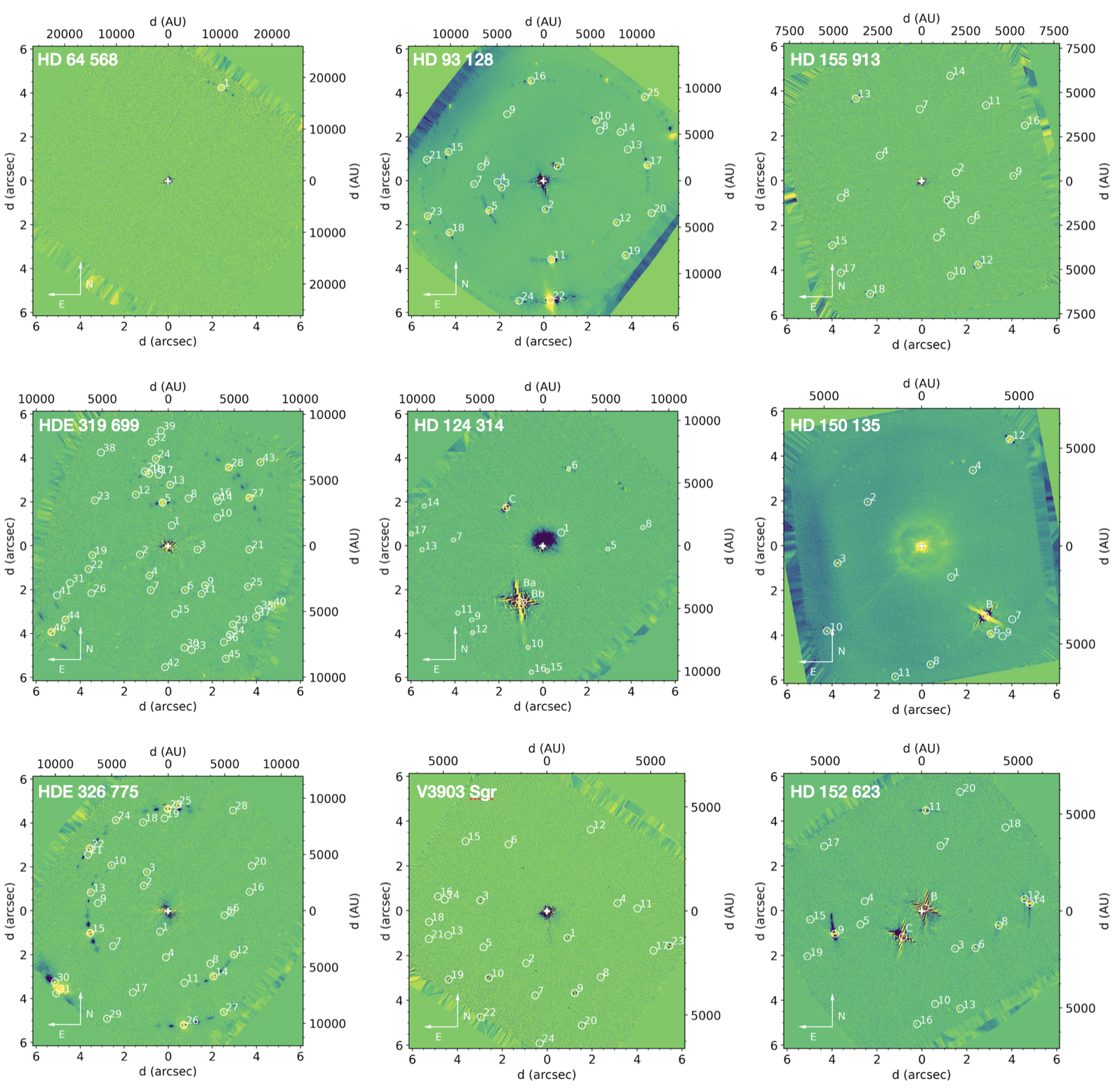}
      \caption{Final post-processed PCA/ADI IRDIS images for the first 9 targets.}
         \label{Fig: final IRDIS images 1}
   \end{figure*}
   
      \begin{figure*}
   \centering
   \includegraphics[angle=-90, width=18cm]{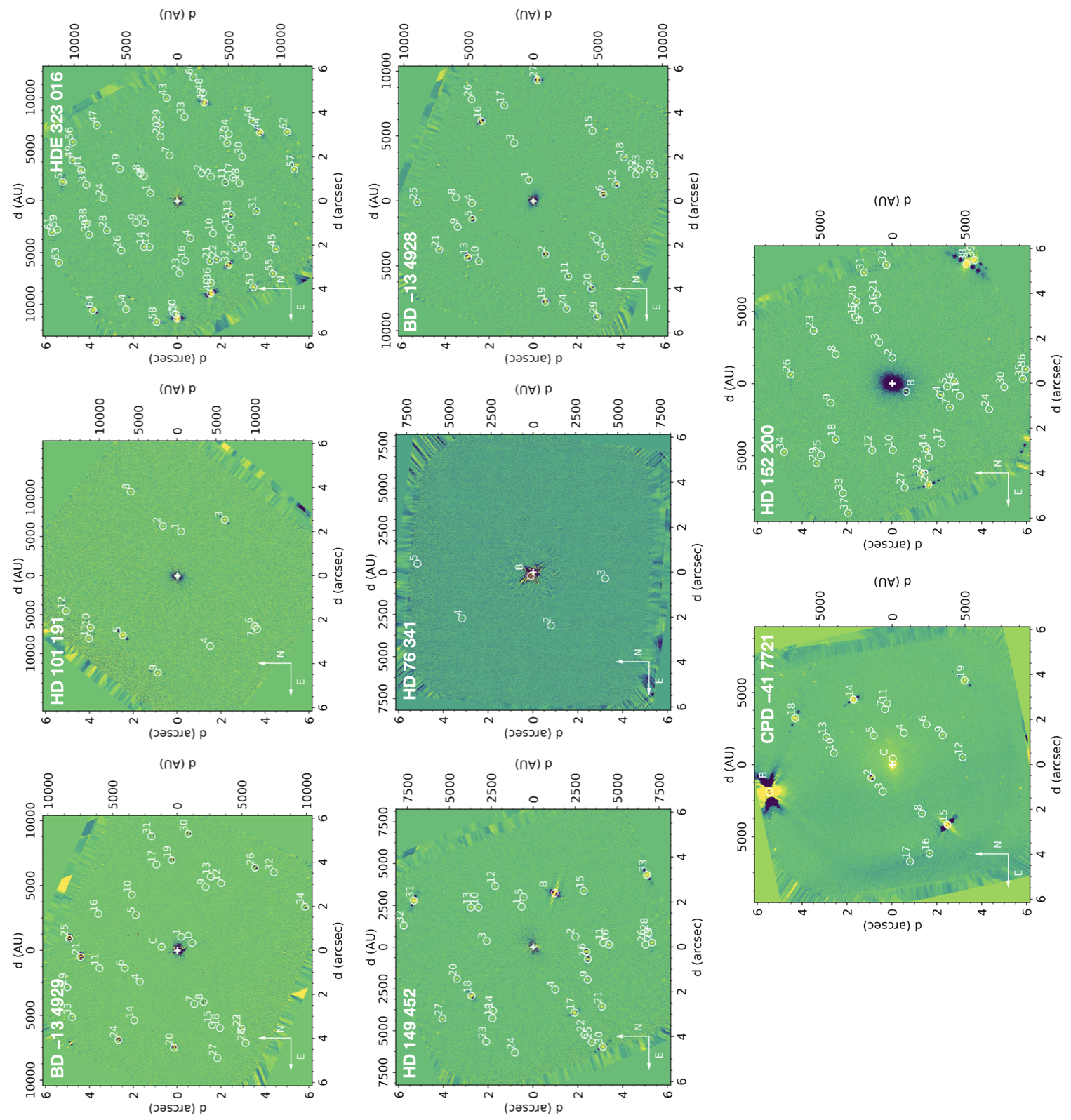}
      \caption{Final post-processed PCA/ADI IRDIS images for the last 8 targets.}
         \label{Fig: final IRDIS images 2}
   \end{figure*}


\section{Results}\label{Sec:results}
The visual inspection of the final IFS and IRDIS PCA images displayed in Figs.~\ref{Fig: final IFS images}, ~\ref{Fig: final IRDIS images 1}, and~\ref{Fig: final IRDIS images 2} reveals the presence of six possible companions in the IFS frames and many other point-like sources in the IRDIS FOVs.  
To evaluate which ones are true detections, we estimated their S/N values using the appropriate function implemented in VIP.  This module computes the S/N at every pixel of an image by measuring the signal in a 1 FWHM diameter aperture and comparing it to the standard deviation of the other resolution elements in an annulus at the same radial distance from the center of the frame. It uses the approach described in \citet{Mawet2014} on small sample statistics, which applies a student t-test to determine the S/N and contrast in high contrast imaging observations. In our study, we adopted a S/N threshold of 5 to assess if a source on the image is a true detection.

\subsection{Source characterization}\label{Sec:source characterization}
Once we identified the true physical objects on our images, we extracted position and flux for each source.
As described by \citet{Rainot2020}, we adopted two different techniques based on the radial separations of the sources. For all IFS sources and all IRDIS source within 2\arcsec, we used a negative fake companion approach, implemented in VIP (see Sect.~\ref{Sec:sources within 2}). For sources with separation beyond 2\arcsec, we adopted a PSF fitting routine \citep{Bodensteiner2020}, as these objects do not suffer from the central star PSF influence in the derotated and collapsed cube.

\subsubsection{Sources within 2\arcsec}\label{Sec:sources within 2}
For all IFS detections and for all sources with angular separations smaller than 2\arcsec~from the central star in the final IRDIS images, we first measured the flux with aperture photometry at each wavelength. We used this initial guess as a starting point of a Simplex Nelder-Mead optimisation implemented in VIP. VIP estimates position and flux for each source by applying a NEGative Fake Companion technique (NEGFC), which consists in inserting negative artificial companions in each individual frame before running PCA, varying at the same time their brightness and location. The residuals are then calculated in the final PCA images and compared to the background noise of all resolution elements in an annulus at the same radial distance. The combination of brightness, separation, and position angle that minimizes the residuals is estimated through a Nelder-Mead minimization algorithm. The artificial companions are obtained from the uncoronagraphic images of the central star. As the Nelder-Mead optimization does not return the uncertainties on the estimates of the parameters, we implemented a set of Monte-Carlo simulations to compute the accuracy of our algorithm. For each wavelength, we inserted 25 artificial sources at the same radial distance and with the same flux as a given detection, but varying their position angles. We  once again measured the flux and location of the injected fake sources using the negative fake companion algorithm and compared them to the true values. The standard deviation of the measurements of each parameter gives us an estimate of the 1$\sigma$ error.  

\subsubsection{Sources beyond 2\arcsec}\label{Sec:sources beyond 2}
Beyond 2\arcsec, the contribution of the central star is negligible and it is the background noise that is dominant (see Sect~ \ref{sec:Sensitivity limits}). Therefore, for point-like objects beyond this separation, the use of ADI and SDI techniques is not necessarily needed to derive precise astrometry and photometry. 
Following \citet{Rainot2020}, we adopted a standard PSF-fitting technique as described in \citet{Bodensteiner2020}, which is based on the \texttt{photutils}\footnote{\url{https://photutils.readthedocs.io}} python package along with an effective PSF model developed by \citet{Anderson2000}.
    
We implemented this method on the derotated and collapsed images for both K1 and K2 and we adopted the IRDIS uncoronagraphic images in each band as accurate PSF models for the fit. The PSF is then fitted to each source individually to estimate the best positions and flux values in K1 and K2, together with their uncertainties. This technique is particularly useful for sources that are too close to the edges of the frames for the NEGFC technique to work.

\subsubsection{Final error budget}
The methods described in Sects. \ref{Sec:sources within 2} and \ref{Sec:sources beyond 2} only take into account the uncertainties related to the algorithm used. For the total errors on the photometry we also accounted for the flux variations of PSFs of the central stars described in Sect.~\ref{Sec:obs and data}. To estimate the total uncertainties on the separation and position angle of each source, we adopted the plate scale and astrometric calibration precision given by \citet{Maire2016} and the ESO SPHERE user manual. The final errors are obtained by a quadratic sum of the algorithm measurement errors (either from the Monte-Carlo simulations or PSF fitting), the star's center position uncertainty \citep[1.2 mas, from][]{Zurlo2016}, the plate scale precision of 0.02 mas/pix for IFS and of 0.021 mas/pix for IRDIS, the dithering procedure accuracy \citep[0.74 mas,][]{Zurlo2016}, and the true north uncertainty ($\pm 0.08\deg$).
The summary of the properties and corresponding uncertainties of all the sources found around each star are available in the online material\footnote{The database is available in electronic form at the CDS via anonymous ftp to cdsarc.u-strasbg.fr (130.79.128.5) or via \url{http://cdsweb.u-strasbg.fr/cgi-bin/qcat?J/A+A/}}.

 \begin{table*}[t]
 \caption[]{\label{table:Fastwind Parameters} Assumed stellar parameters for the FASTWIND calculation of the calibrated spectra. Effective temperatures T$_{eff}$, logarithmic luminosities log L, masses M, and radii R are taken from \citet{Martins2005} and \citet{Trundle2007}, for O- and B-type stars, respectively. The mass-loss rate ($\dot{M}$) and terminal wind velocities ($v_\infty$) are computed following the prescriptions given by \citet{Vink2001}.}
\begin{tabular}{lccccccc}
 \hline \hline
  Object ID &
  SpT &
$T_{\mathrm{eff}}$ [K] & $\log L$(\lsol) & $M$ [\Msun] & $R$ [\rsun] & $\dot{M}$ [\Msun/yr] &
  $v_\infty$ [km/s] \\
 \hline
HD~\num{93128}   & O3.5~V((fc))z &  43854  &  5.75  & 52.44  &  13.11 & 3.24e-06 & 3212.75\\
HDE~\num{319699}   & O5~V((fc)) &  40862    &  5.49  & 38.08  &  11.20 & 1.29e-06 & 2962.01\\
HD~\num{124314}   & O6~IV(n)((f)) &  38867   &  5.32  & 30.98  &  10.11 & 6.82e-07 & 2811.98\\
HD~\num{150135}   & O6.5~V((f))z  &  37870  &  5.23  & 28.00  &  9.61 & 4.77e-07 & 2741.98 \\
HD~\num{150135}   &  O8: &  34877   &  4.96  & 20.76  &  8.29 & 1.55e-07 & 2542.05 \\
HDE~\num{326775}   & O6.5~V(n)((f))z &  37870    &  5.23  & 28.00  &  9.61 & 4.77e-07 & 2741.98 \\
V3903~Sgr   & O7~V(n)z &  36872    &  5.14  & 25.29  &  9.15 & 3.32e-07 & 2670.61\\
V3903~Sgr   & B0:~V &  32020  &  4.31  & 15.00  &  5.10 & 7.12e-09 & 2755.87\\
HD~\num{152623}   & O7~V(n)((f)) & 36872  &  5.14  & 25.29  &  9.15 & 3.32e-07 & 2670.61 \\
BD~$-$13\degr~4929   & O8~V &  34877   &  4.96  & 20.76  &  8.29 & 1.55e-07 & 2542.05\\
BD~$-$13\degr~4929   & B0.5:~V &  29700  &  5.14  & 16.00  &  5.67 & 2.55e-08 & 2721.23 \\
HD~\num{149452}   & O9~IVn &  32882  &  4.77  & 17.08  &  7.53 & 6.63e-08 & 2419.32 \\
HD~\num{76341}   &  O9.2~IV &   32383  & 4.72 & 16.31  &  7.35 & 5.24e-08 & 2392.93  \\
BD~$-$13\degr~4928   & O9.5 V &   31884  & 4.68 & 15.55  &  7.18 & 4.35e-08 & 2364.01 \\
CPD~$-$41\degr~7721 & O9.7~V:(n) &  31884  &  4.68  & 15.55  &  7.18 & 4.35e-08 & 2364.01 \\
HD~\num{152200}   & O9.7~IV(n) &  31385    &  4.64  & 14.79  &  7.00 & 3.61e-08 & 2334.97\\
\hline
\end{tabular}
\end{table*}

\subsection{Spurious association probabilities}
To identify which of the detected sources could be bound companions, we estimated on a statistical base the probability of chance alignment association. 
Following \citet{Rainot2020}, we defined the probability of spurious association ($P_\mathrm{spur}(\rho_i|\Sigma(K_i))$) as the probability that at least one object is found by chance at an angular separation from the central star $\rho$ smaller or equal than that of the $i$th companion (i.e., $\rho \le \rho_i$), given the local source density $\Sigma$ of stars at least as bright as $i$ ($K \le K_i$). 
To compute $P_\mathrm{spur}$, we first evaluated the local field density $\Sigma(K_i)=N_\mathrm{obj}(K \le K_i)/(\pi r^2)$ of objects at least as bright as the companion $i$ in a $\pi r^2$ surface. To do so, we used the Gaia \citep{Gaia2018} DR2 catalog to evaluate the number and brightness of companions in a $r$=2\arcmin~radius region around each target.  To convert the Gaia magnitudes into $K$-band magnitudes, we used the color relations given by \citet{Evans2018}. For each $i$ source, we then used a Monte Carlo approach to generate 10,000 populations of $N_\mathrm{obj}(K \le K_i)$ stars uniformly distributed in $\pi r^2$. The probability of spurious association is thus obtained as the fraction of Monte Carlo simulations in which at least one star is found at the separation $\rho \le \rho_i$. 

All properties and probabilities for sources with $P_\mathrm{spur} <$ 5\% are given in Table~\ref{table:Properties 1}. 
For the objects with low spurious association probability (e.g., $P_\mathrm{spur}<0.05$), which are most likely bound companions, it is essential to obtain a confirmation of common proper motions and a characterization of orbital motion in the future to definitively demonstrate a true physical association.

 \begin{table*}[t]
 \centering
 \caption[]{\label{table:Properties 1} Angular separations ($\rho$), position angles (PA), magnitude contrasts in the $K$1 and $K$2 bands ($\Delta$K12), and spurious alignment probabilities ($P_\mathrm{spur}$) for the sources with $P_\mathrm{spur} <$ 5\% around our targets. The object numbers correspond to those in Figs.~\ref{Fig: final IRDIS images 1} and ~\ref{Fig: final IRDIS images 2}.}
\begin{tabular}{lccccccccccc}
 \hline \hline
  Object & $\rho$ (\arcsec) & $\sigma$ $\rho$ (\arcsec) & PA ($\degr$) & $\sigma$ PA ($\degr$) & $\Delta$K1 & $\sigma$ $\Delta$K1 & $\Delta$K2 &  $\sigma$ $\Delta$K2 & $P_\mathrm{spur}$ & M [\Msun] & age [Myr]\\
 \hline
HD~\num{64568}  &  &  &  &  &   &   & \\
HD~\num{93128}   &  &  &  &  &   &   & \\
   1 & 0.926 &     0.002 & 319.03 &       0.08 &         7.91 &             0.18 &         7.55 &             0.12 &    0.02 & 0.25-0.4  & 0.3-1.5\\
   2 & 1.286 &     0.003 & 182.94 &       0.08 &         8.04 &             0.18 &         7.70 &             0.21 &    0.04  & 0.2-0.4  & 0.3-1.5 \\
HD~\num{155913}  &  &  &  &  &   &   & \\
HDE~\num{319699}  &  &  &  &  &    &   & \\
   1 & 0.945 &     0.011 & 350.91 &       0.44 &        11.94 &             0.28 &        11.25 &             0.50 &    0.01   &  -- & --\\
   2 & 1.340     &     0.006 & 107.00 &  0.13 &       10.80      &             0.20 &        10.80        &             0.25 &    0.03   &  -- & -- \\
   3 & 1.348 &     0.005 & 263.21 &       0.08 &         9.87 &             0.05 &         9.91 &             0.09 &    0.02   &  -- & --\\
   4 & 1.590 &     0.003 & 148.00 &       0.08 &         9.87 &             0.03 &         9.91 &             0.12 &    0.02   &  -- & --\\
   5 & 1.996 &     0.004 &   7.25 &       0.08 &         7.31 &             0.01 &         7.19 &             0.02 &    0.01  &  0.3-0.4 & 1.3-1.4 \\
   6 & 2.150 &     0.004 & 201.04 &       0.08 &         9.59 &             0.01 &         9.51 &             0.03 &    0.05   & 0.25  & $>$8\\
HD~\num{124314}  &  &  &  &  &    &   & \\
   1 & 1.010 &     0.014 &  306.4        &       0.3 &         11.4      &             0.2 &         11.9 &              0.2 &    0.025 &  & --  \\
   C & 2.413 &     0.004 &  43.55 &       0.08 &         6.079 &             0.003 &         6.002 &             0.003 &    0.007 &  1-1.6 & 1.7-4.9  \\
   Ba & 2.708 &     0.005 & 156.66 &       0.08 &         1.946 &             0.005 &         1.901 &             0.005 &    0.001  &  5-7  & 0.1-0.5  \\
    &   &     &  &   &     &    &    &    &     &  15-20  &  1.5-5.2 \\
   Bb & 2.722 &     0.006 & 161.00 &       0.12 &         4.49 &             0.13 &         4.53 &             0.14 &    0.003   & 3-3.5  & 1.7-2.4 \\
HD~\num{150135}   &  &  &  & &    &   & \\
   1 & 1.918 &     0.004 & 223.45 &       0.08 &         9.55    &             0.02 &         10.7        &      1.7   &    0.04   &  0.1-0.13  &  3.5-4.5\\
   3 & 3.820 &     0.007 & 101.57 &       0.08 &         7.59    &             0.01 &         7.48 &             0.01 &    0.05   &  0.6-0.8  &  2.7-4.3 \\
   B & 4.209 &     0.007 & 222.30 &       0.08 &         2.54    &             0.01 &         2.50        &             0.01 &    0.002   & 5  & 0.5 \\
      &   &   &   &   &    &    &        &    &  &  12  &  5.4-5.5\\
HDE~\num{326775} &  &  &  &   &   &   & \\
   1 & 1.019 &     0.011 & 159.03 &       0.30 &        11.35 &             0.18 &        10.98 &             0.40 &    0.02   &  -- &  --\\
   2 & 1.592 &     0.004 &  43.88 &       0.09 &        10.27 &             0.04 &         9.87 &             0.18 &    0.05   & 0.1  & 7.3-7.9\\
  15 & 3.668 &     0.006 & 105.89 &       0.08 &         6.33 &             0.01 &         6.21 &             0.07 &    0.05   &  0.5-1.4 & 1.1-5.4\\
  31 & 6.292 &     0.011 & 126.57 &       0.08 &         5.00 &             0.015 &         4.30 &             0.01 &    0.04   & --  & --\\
V3903~Sgr   &  &  &  &  &    &   & \\
   1 & 1.519 &     0.003 & 217.58 &       0.08 &         7.93 &             0.02 &         8.02 &             0.05 &    0.01   &  -- & --\\
HD~\num{152623}  &  &  &  &  &  &   &  \\
   B & 0.212 &     0.002 & 311.62 &       0.13 &         1.56 &             0.03 &         1.99 &             0.13 &    0.000   & 6  & 0.3\\
   C & 1.452 &     0.003 & 143.09 &       0.08 &         2.958 &             0.004 &         2.831 &             0.004 &    0.000   &  3.5 & 1.5\\
   9 & 4.023 &     0.007 & 104.69 &       0.08 &         6.045 &             0.004 &         5.967 &             0.004 &    0.04   &  0.9-1.5  & 1.8-4.4\\
BD~$-$13\degr~4929  &  &  &  &  &   &   & \\
       1 & 0.652 &     0.010 & 249.4  &       0.9 &          9.5 &              0.5 &        12.6 &            0.9 &    0.01 &  0.1-0.2 &  0.8-5.\\
       C & 0.732 &     0.010 & 345.7  &       0.2 &          7.7 &              0.3 &         9.3 &             0.9 &    0.008  &  0.2-0.6 &  1.3-5.\\
      D & 0.761 &     0.005 & 207.5  &       0.1 &          7.50 &              0.04 &         7.3 &             0.4 &    0.008  &  0.2-0.8 &  1.2-9.\\
HD~\num{101191}   &  &  &  &  &   &   & \\
HDE~\num{323016}   &  &  &  &  &   &   & \\
HD~\num{149452}  &  &  &  &  &    &   & \\
   B & 2.596 &     0.005 & 248.27 &       0.08 &         4.359 &             0.003 &         3.410 &             0.003 &    0.005    & 1.8-3  &  2.2-8. \\
      &  &      &  &        &    &       &   &   &   &   3 & 2.1-6.4\\
HD~\num{76341}   &  &  &  &  &   &   & \\
   B & 0.159 &     0.002 &   59.67 &       0.32 &        3.72 &             0.04 &        3.47 &            0.21 &    0.00 & 2.7-4   & 2-6.8 \\
   2 & 1.053 &     0.003 &  183.52 &       0.10 &        10.77 &            0.03 &        11.93 &            1.81 &    0.006  &  -- & -- \\
   3 & 2.469 &     0.005 & 108.14 &       0.09 &        12.77 &            0.06 &        12.09 &            0.06 &    0.04 &  -- &  -- \\
BD~$-$13\degr~4928   &  &  &  & &    &   & \\
  1 & 0.990 &     0.002 & 280.79 &       0.08 &         7.92 &             0.01 &         7.75 &             0.03 &    0.02  & 0.1  & 2.2-2.4 \\
CPD~$-$41\degr~7721  &  &  &  &  &   &   & \\
   C & 0.267 &     0.003 & 263.22 &       0.14 &         7.40 &             0.03 &         7.15 &             0.44 &    0.001 & 0.16-0.3 & 1.-4.8\\
   2 & 1.085 &     0.002 &  31.928 &       0.08 &         6.16 &             0.02 &         6.10 &             0.42 &    0.01  & 0.6-0.8  &  2.7-4.3\\
  15 & 3.638 &     0.006 & 132.73 &       0.08 &         3.644 &             0.002 &         3.583 &             0.002 &    0.05  & 2.2-3.5 &  2.6-5.4\\
  B & 5.616 &     0.009 &  12.55 &       0.08 &         1.440 &             0.003 &         1.401 &             0.003 &    0.04  & 5.0  & 6.2-6.3  \\
      &   &       &    &         &     &  &           &    &  &  12-34 &  0.6-2.5 \\
HD~\num{152200}  &  &  &  &  &   &   & \\
   B & 0.729 &     0.003 & 150.65 &       0.10 &         8.80 &             0.13 &         8.57 &             0.13 &    0.01   &  0.1 & 4.9-5.7\\
   2 & 1.178 &     0.005 & 270.00 &       0.18 &        10.42 &             0.10 &        11.07 &             0.76 &    0.05   &  --  &  --\\
\hline
\end{tabular}
\end{table*}

\subsection{Absolute flux values}
 Knowing the flux calibrated spectrum of the central star in the wavelength range covered by our SPHERE observations ($Y$ to $K$) is required to compute the absolute fluxes of possible bound companions in the images. In fact, under the assumption that the
same extinction affects both the primary and its companions, the unreddened primary spectral energy distribution allows us to derive the companions spectral energy distribution and thus characterize them through a comparison with atmosphere models (see Sect.~\ref{Results:fitting} ).  As such a spectrum is not easily available, we modelled the spectral energy distribution of the targets with likely bound sources (those with $P_\mathrm{spur} <$ 5\%) with the non-local thermodynamic equilibrium (non-LTE) atmosphere code FASTWIND \citep[][]{Puls2005}.  When the central object is composed by a spectroscopic multiple system, each component is modelled separately at first, and then combined to obtain a unique spectrum. This is the case for HD~\num{150135}, BD~$-$13\degr~4929, and V3903~Sgr. The assumed input parameters for FASTWIND for each component of the central object are presented in Table~\ref{table:Fastwind Parameters}.  Since we only characterized sources with $P_\mathrm{spur} <$ 5\%,  Table~\ref{table:Fastwind Parameters} only includes the stars hosting likely bound companions. The parameters for the computation were based on the spectral type characterization found in the literature (see Table \ref{Sect:sample}) and the observational O-star calibration tables from \citet{Martins2005}. Parameters for the B star components were instead found in \citet{Trundle2007}. As inputs for FASTWIND, we  calculated the mass-loss rate ($\dot{M}$) and terminal wind velocities ($v_\infty$) for each star following \citet{Vink2001}. 
We remark that in this process and later on in Sect.~\ref{Results:summary}, it is necessary to adopt a reference radius for the sphere at the surface of which the flux is computed. Without loss of generality, we arbitrarily adopted a value of 100~\rsun, although this value has no physical meaning, or impact in our calculation. 

\subsection{Spectral fitting}\label{Results:fitting}
Similarly to what was done by \cite{Rainot2020} for QZ Car, we used the low-resolution IFS spectrum of all IFS detections to constrain their stellar parameters. 
To do so, we used both the ATLAS9 LTE atmosphere models \citep{Castelli2003}, covering the 3500-50000 K temperature range, and the LTE PHOENIX models \citep[2300-12000 K,][]{Husser2013} for $T$ $<$ 3500 K. To each age value in the pre-main sequence (PMS) evolutionary tracks of \cite{Siess2000} below 7\Msun~or of \cite{Brott2011} above, we associated an atmospheric model and we quantitatively compared it to the flux calibrated SED of each detection.
For the comparison, we rebinned each model to the 39 IFS wavelength channels and the 2 IRDIS bands and we estimated the corresponding $\chi^2$ by taking into account the uncertainty on the flux calibrated spectrum.

Several combinations of stellar parameters are consistent with the observations.
In the next section (Sect. \ref{Results:summary}), we summarize (object by object) the results of the SED fitting procedure and the best fit parameters corresponding to the 95\% confidence interval.

Our IRDIS observations provide us with only two independent wavelength channels, K1 and K2, at 2.110 and 2.251~$\mu$m, respectively. This does not allow us to constrain the shape of the SED, however, it does offer the possibility to assess the object absolute $K$-band magnitude under the assumption that the IRDIS sources are located at the same distance as the central star. Therefore, for all IRDIS detections with $P_\mathrm{spur} <$ 5\%, which are most likely to be bound companions, we compared the K1 and K2 absolute fluxes to the ATLAS9 and PHOENIX models.  For many of them, we were able to find solutions that are in agreement with the central star age (as reported in Table~\ref{Table:O star properties}). For those stars, the ranges of masses and ages that are consistent with the IRDIS observations are given in Table~\ref{table:Properties 1}. For several objects, we could not find a good fit within the given age range, possibly indicating that they do not have physical companions in common. Finally, given our grid of models, we are also limited to PMS stars more massive than 0.1~M$_\odot$. Some of the very faint sources for which we could not find a proper fit could be shown to be objects at the stellar-substellar boundary that are not covered by our models.

\subsection{Summary}\label{Results:summary}

\begin{itemize}
    \item HD~\num{64568} has been associated with an O3~V((f*))z SpT \citep{Sota2014}. It has recently been classified as runaway from the southern component of NGC 2467 by \cite{MaizApellaniz2020} and \cite{MaizApellaniz2021}. We did not find close companions in the IFS field nor in the IRDIS larger FOV. This supports the idea that all O stars are born in multiple systems.\\
      
    \item HD~\num{93128} (Trumpler 14 2) is part of Trumpler 14 and has been classified as O3.5~V((fc))z SpT  \citep{Sota2014}. We do not detected any IFS companion. Two IRDIS sources are however found with $P_\mathrm{spur} \leq 5$ \% at separations of 0\farcs93 and 1\farcs29 and for which the fits of the $K$-band magnitudes is consistent with masses of 0.25-0.4 and 0.2-0.4 \Msun{}, respectively, with ages between 0.3-1.5 Myrs. A more detailed analysis of the system, including the reanalysis of the B visual companion (classified as B0.2~V by \cite{MaizApellaniz2021}) will be presented in \cite{Rainot2021}. \\
      
    \item HD~\num{155913} is a classified as a possible runaway given that the proper motion points away from the young stellar cluster NGC 6822, which is located half a degree away from the star \citep{MaizApellaniz2018}. HD~\num{155913} is a O4.5~Vn((f)) SpT according to the GOSS catalog \citep{Sota2014}. However, it is reported as SB2 in the OWN data \citep{Barba2010,Barba2017}, so the width observed in the GOSS spectrograph could be due to unresolved orbital motion. A visual companion is reported by \citet{Aldoretta2015}. No relevant sources are found in our IFS and IRDIS images.\\
    
    \item HDE~\num{319699} is an O5~V((fc)) type star \citep{Sota2014} and it is part of NGC 6334 \citep{Russeil2017}. According to our IRDIS images, the star is part of a crowded field, but no close IFS companions or bright IRDIS sources are present in the data. Six sources within roughly 2\arcsec~are found with $P_\mathrm{spur} \leq 5$ \%.\\

    \item HD~\num{124314} is a O6~IV(n)((f)) Aa-Ab binary with a separation of 1.5 mas, separated by the O9.2~IV(n) visual components Ba-Bb  by 2\farcs5 \citep{Sota2014}. A third visual component C was found by SMaSH+ at 2\farcs8.  No additional companions are seen in our IFS observations, but all visual companions are also detected in our IRDIS images with $P_\mathrm{spur} \lesssim 1$\%. According to our fit, all visual components could be coeval with the central star system.  \\
    
    \item HD~\num{150135}: it has been recently classified as SB2 by \cite{MaizApellaniz2020}, composed by a O6.5 V((f))z primary and an O8 secondary component. SMaSH+ found it to be a Aa-Ab+B multiple system. The A-B components are separated by 4\farcs27, whereas Aa-Ab by 0.95 mas. Although we do not detect any companion in the IFS field, the elongated IFS PSF suggests the close inner binary could be currently at a larger angular separation (with a tentative orientation of $\sim 45\degr$). Our IRDIS images show the presence of the B components, which appears to be a possible double system with source 6 ($P_\mathrm{spur}\sim 7$\%). \\
    
    \item HDE~\num{326775} \citep[O6.5~V(n)((f))z,][]{MaizApellaniz2016} is part of RCW 113 HII region \citep{Arias2016}, situated in the southern part of the Sco OB1 association. We did not detect companions in the IFS data. Among the four sources with $P_\mathrm{spur} \leq 5$\% in the IRDIS field, only two of them show $K$-band magnitudes that could be fitted with models in the age range of the central star.    \\

    \item V3903~Sgr is a O7~V(n)z + B0:~V  binary system and it is part of the Sgr OB1 association \citep{Sota2014}. We found a most likely (spurious association probability of 1\%) visual companion at a  separation of 1\farcs5 and with a $\Delta K$ = 8.0 mag. Unfortunately, we cannot obtain an acceptable fit of the $K$-band magnitudes with an age that is consistent with the central star.   \\
    
    \item HD~\num{152623} is a known colliding-wind binary \citep{DeBecker2013}, and a possible runaway \citep[][and references therein]{MaizApellaniz2018}. The components B and C were detected by SMaSH+ at a separation of 0\farcs25 and 1\farcs5, respectively.  HD~\num{152623}~B is detected in both IFS and IRDIS images and our best fit for the spectrum is obtained with a mass of 6 \Msun~at an age of 0.3 Myrs (see Fig.~\ref{Fig: HD152623B spectrum}). HD~\num{152623}~C is detected with a $\Delta$K = 6.13, and it is consistent with being a 3.5 \msun~with an age of 1.5 Myr. Source 9 has also a probability of $P_\mathrm{spur}= 4$\%, suggesting it could be a bound companion to the system as well. It appears to be roughly coeval with other components and  it can be fitted best with a 0.9 \msun~star in the age range of 1.8-4.5 Myrs\\
      
 \begin{figure}
   \centering
   \includegraphics[width=\columnwidth]{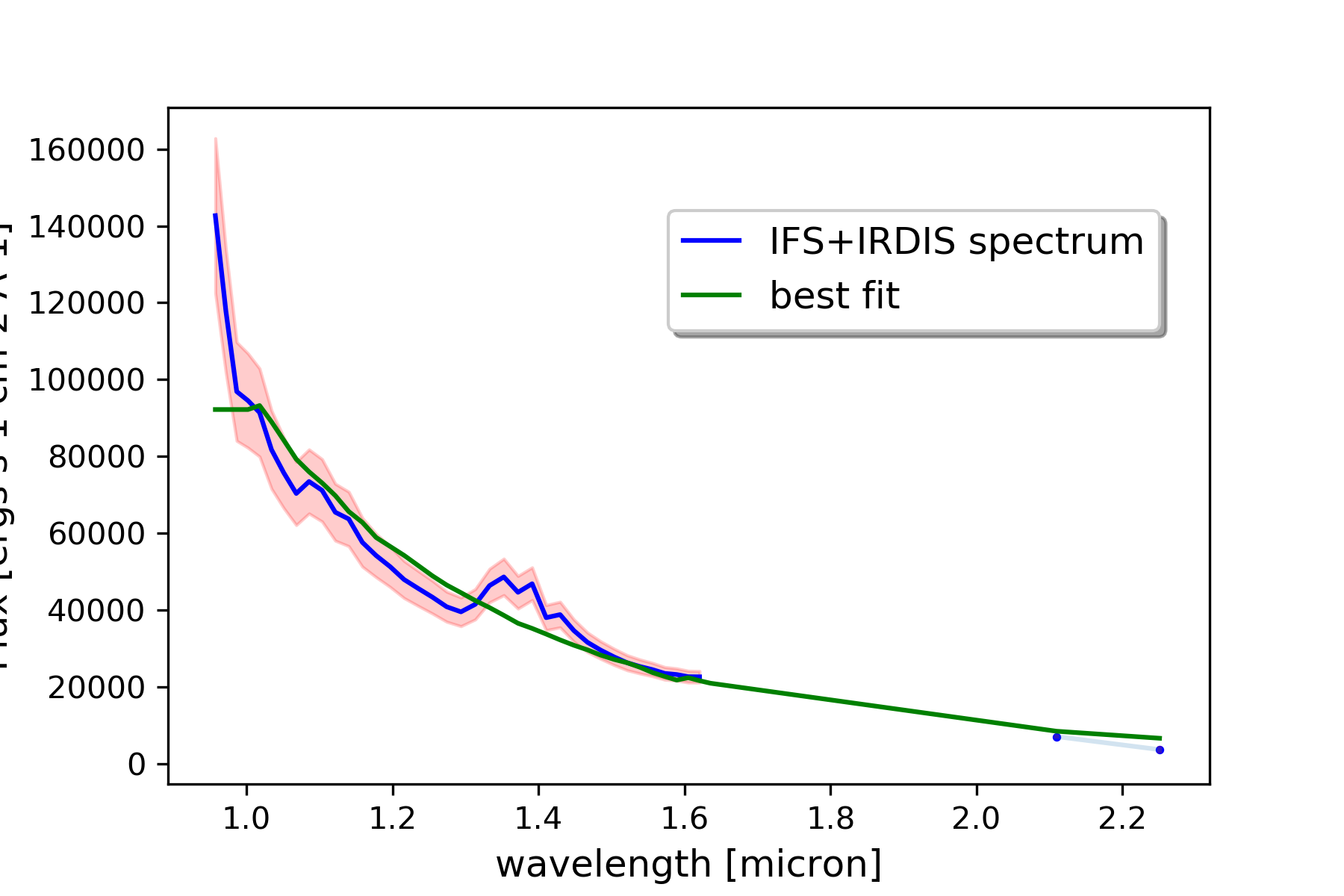}
    \caption{Observed spectrum for HD~\num{152623}~B. }
    \label{Fig: HD152623B spectrum}
\end{figure}

    \item BD~$-$13\degr~4929 it is part of NGC 6611 and it is found to be a SB3 by \citep{Sana2009}. The triple system is composed by an O8~V and two B0.5:~V+B0.5:~V SpT stars \citep{MaizApellaniz2021}.
    We detected two companions in IFS, which are also confirmed by the IRDIS images. According to isochrones fitting, they are both estimated to be 0.2 \Msun~with a best age value of 1.8 and 1.9 Myrs, respectively, in good agreement with recent age determinations of NGC6611 \citep{Bonatto2006, Martayan2008}. The spectra of the companions are presented in Figs.~\ref{Fig: BD-134929C spectrum} and ~\ref{Fig: BD-134929D spectrum}. In IRDIS we also detect a fainter source at 0\farcs65, which is not visible in IFS.  The $K$-band fluxes are consistent with a 0.1 \msun~star with an age in the range of 0.8-5 Myrs.  \\ 
   
   \begin{figure}
   \centering
   \includegraphics[width=\columnwidth]{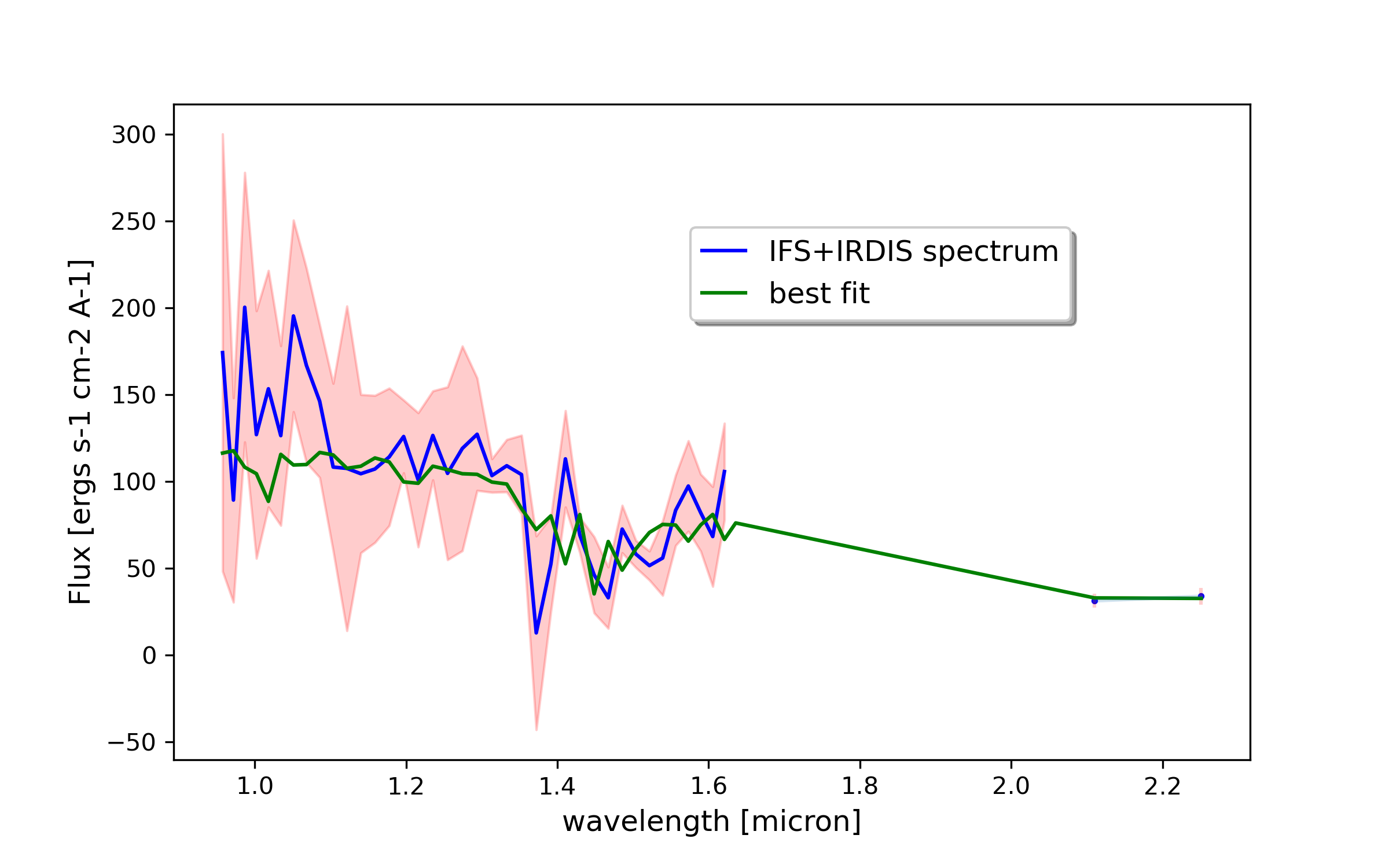}
   \caption{Observed spectrum for BD~$-$13\degr~4929~C. }
   \label{Fig: BD-134929C spectrum}
   \end{figure}
   \begin{figure}
   \centering
   \includegraphics[width=\columnwidth]{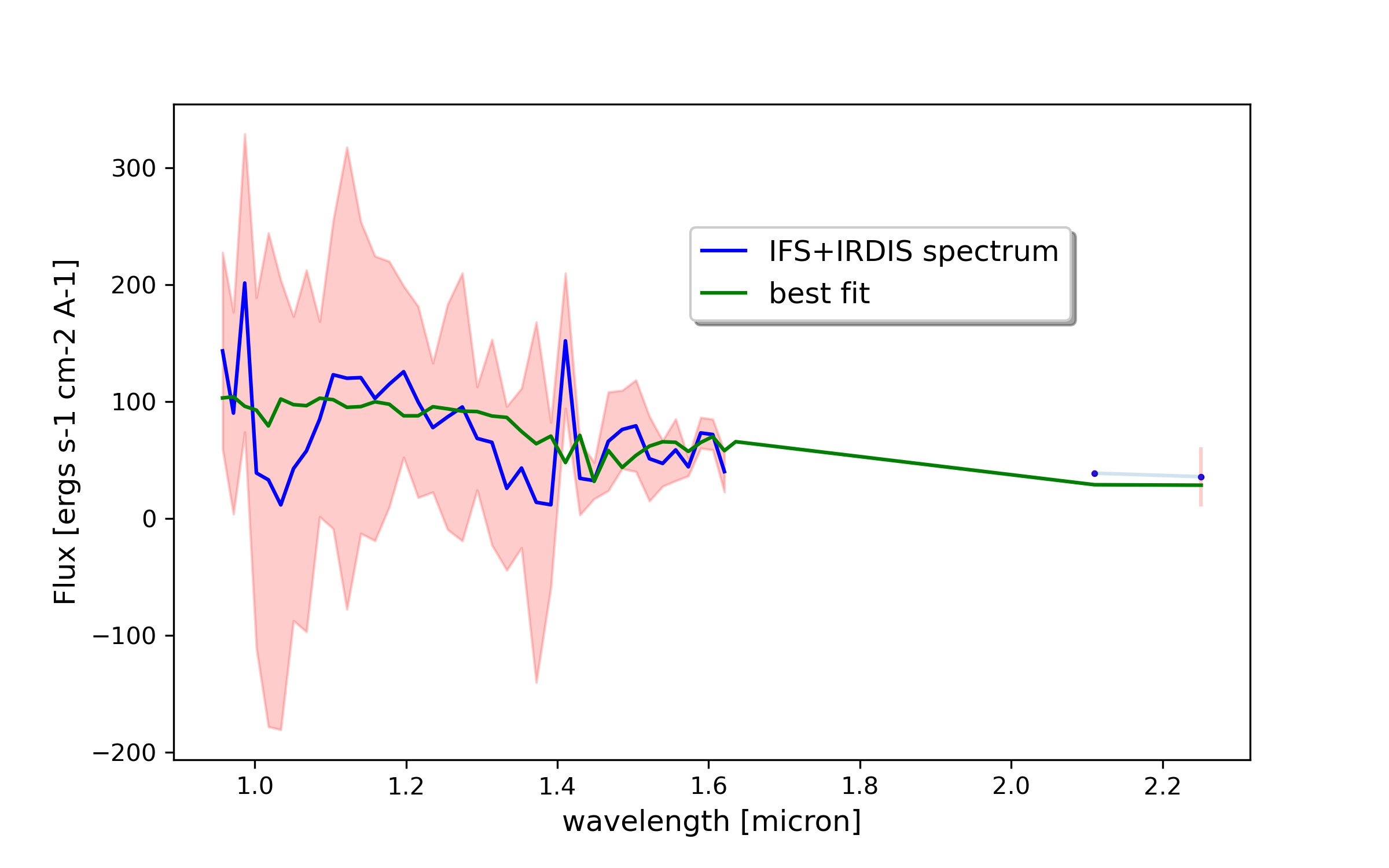}
    \caption{Observed spectrum for BD~$-$13\degr~4929~D. }
    \label{Fig: BD-134929D spectrum}
\end{figure}

    \item HD~\num{101191} \citep[O8~V SpT,][]{Sota2014} is a long-period SB1 system according to \citet{Sana2011} and \citet{Chini2012}. No likely companion appears in IFS nor in IRDIS.  \\
    
    \item HDE~\num{323016} is classified as a O8.5~V SpT and it is part of the S5 HII region according to \citet{Neckel1981}. Similarly to HDE~\num{319699}, the IRDIS FoV reveals a densely populated region around the star, but none of the sources appears to have a $P_\mathrm{spur} \leq 5$\%. \\
 
    \item HD~\num{149452} (O9~IVn) is a relatively isolated O type star \citep{Sota2014}. \citet{Sana2014} reported the detection of a source at 2\farcs7 from HD~\num{149452}. Given the non-detection in their $H$ band image, suggesting a strongly reddened object, they indicated the possibility of it being a background object. In our IRDIS observations, we found a $\Delta$K =3.9 mag source at 2\farcs6, with a $P_\mathrm{spur}= 5$\%. Our $K$-band colors are consistent with 3-28\msun~object with an age of 2-5 Myrs, depending on the evolutionary models used. \\
    
    \item HD~\num{76341} is classified as O9.2~IV by \citet{Sota2014}.  One companion has been detected by SMASH+ \citep{Sana2014} at $\rho$= 169 mas and with a contrast of 3.7 mag in the $H$ band.  \citet{Sota2014} also observed variability in the spectrum of HD~\num{76341}, indicating a possible hierarchical triple system. We re-detected the companion at 159 mas in both IFS and IRDIS observations and we characterized it as a 3.5\msun~object with an age of about 2.4 Myrs. The observed spectrum of the companion is shown in Fig.~\ref{Fig: HD76341B spectrum}. \\
      
\begin{figure}
   \centering
   \includegraphics[width=\columnwidth]{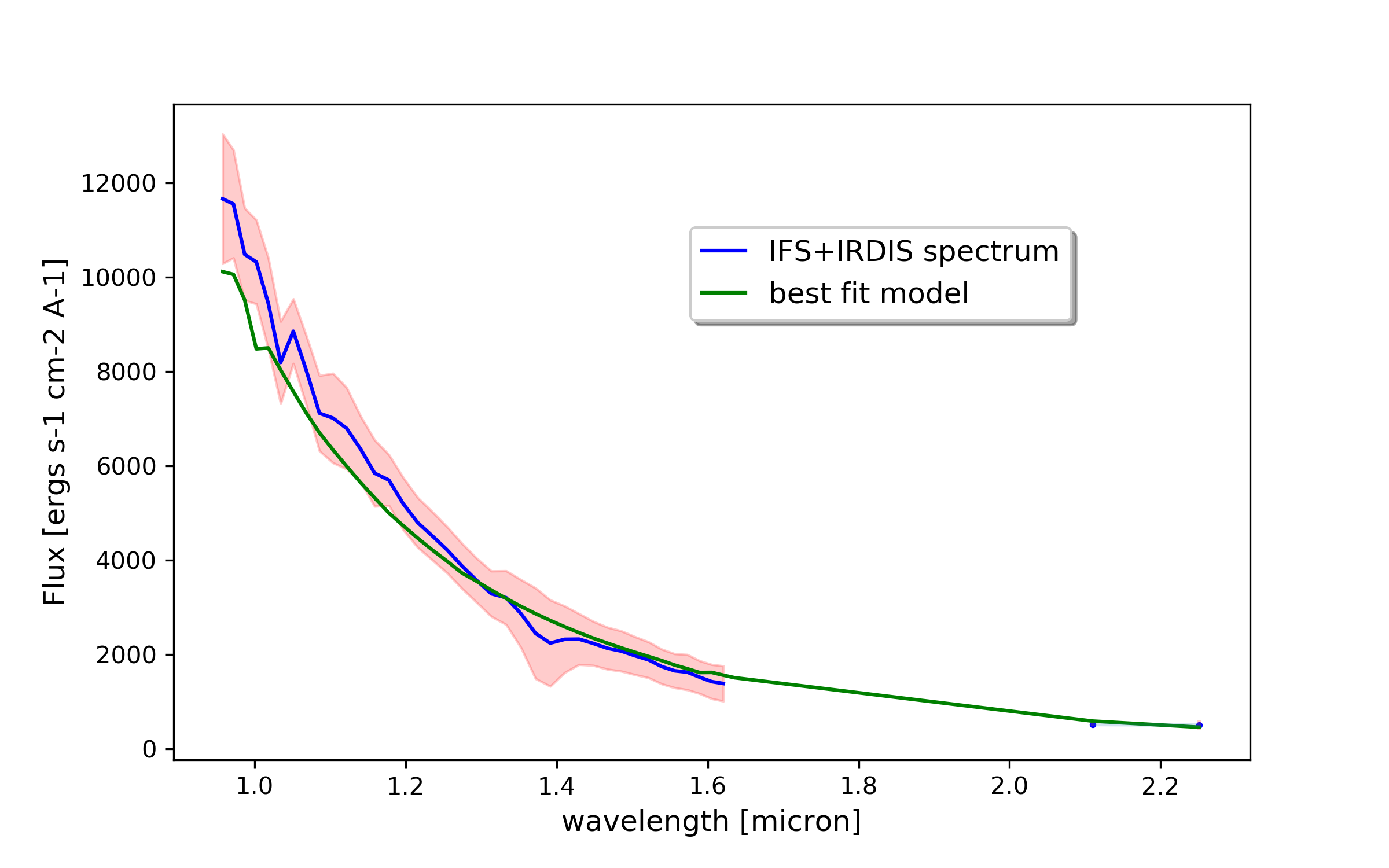}
      \caption{Observed spectrum for HD~\num{76341}B. }
         \label{Fig: HD76341B spectrum}
   \end{figure}
      
    \item BD~$-$13\degr~4928 is a O9.5 V star according to \cite{Sana2009} and a fast rotator (private communication).  We did not find close companions in the  IFS  field. A source at a separation of 0\farcs99 is present in the IRDIS image. It has a $\Delta K$ magnitude of 7.9 and a probability of spurious association of $<2\%$. The best fit mass of this object is 0.1 \msun ~at 2.4 Myr. \\
    
    \item CPD~$-$41\degr~7721 is part of NGC 6321, in the core of the Sco OB1 association, at about 1.5 kpc \citep{MaizApellaniz2021}. According to \citet{MaizApellaniz2016} it is a O9.7~V:(n) star. Together with the  B1.5 V star CPD~$-$41\degr~7721B, it forms a visual double star with the two components separated by 5\farcs8. Besides confirming the already known visual companion with the IRDIS observations, we detected a new fainter and closer companion in IFS. We classified it as a 0.16 \Msun~star, with an age of 3.2 Myrs. The spectrum of CPD~$-$41\degr~7721~C is presented in Fig.~\ref{Fig: CPD-417721B spectrum}.  Another IRDIS source with a $\Delta$K = 6.13 is detected at a separation of  1\arcsec~with a $P_\mathrm{spur}\sim 1$\%, making it a likely bound companion. The best fit for this object is obtained with a best fit mass of 0.7 \msun~and an age of 3.8 Myrs. \\
      
\begin{figure}
   \centering
   \includegraphics[width=\columnwidth]{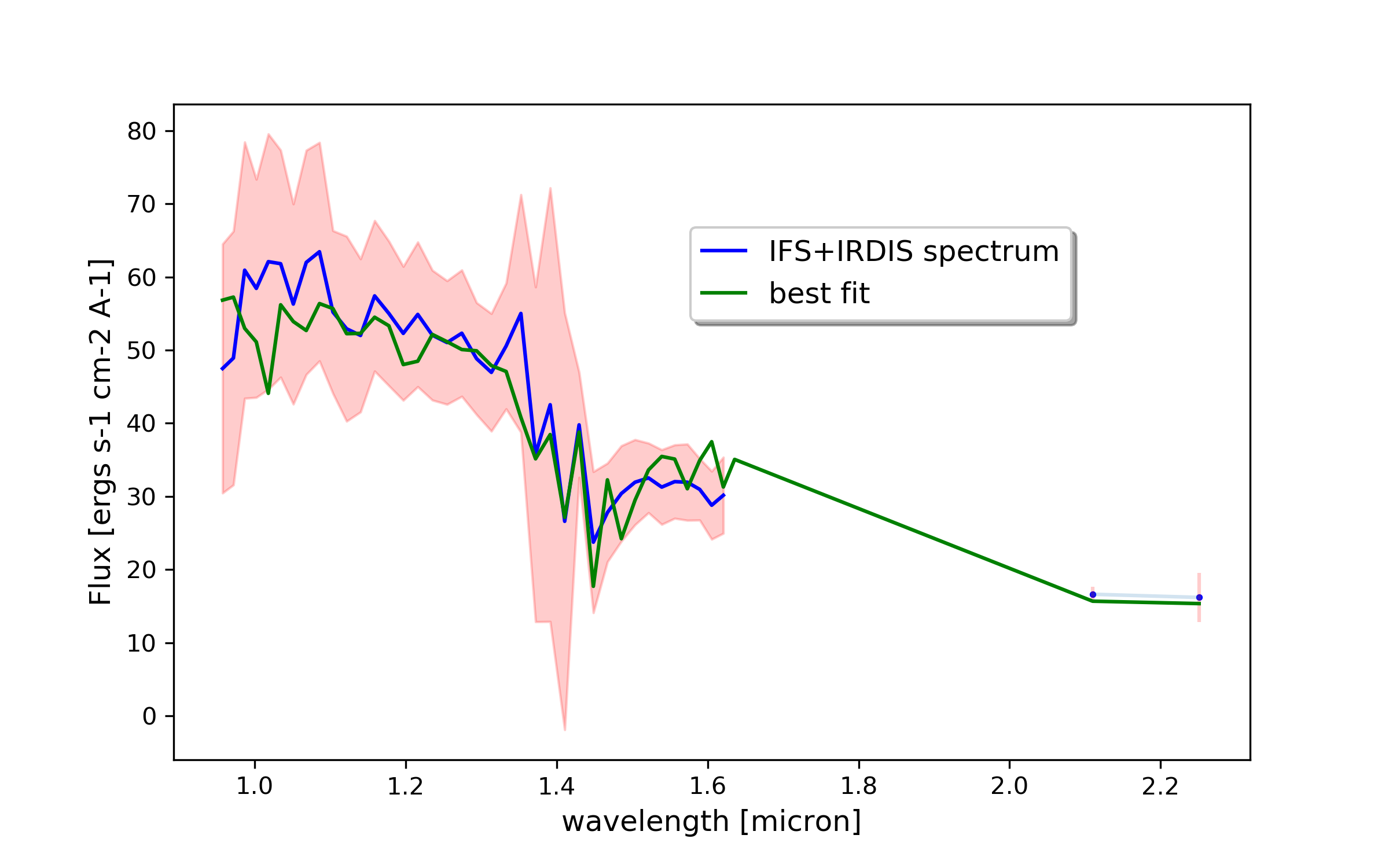}
      \caption{Observed spectrum for CPD~$-$41\degr~7721~C. }
         \label{Fig: CPD-417721B spectrum}
\end{figure}
      
    \item HD~\num{123056} is a field star and it is found to be a hierarchical triple system by \citet{Mayer2017}. The PSF of IFS  looks elongated, also indicating the multiple nature of the central object. No further IFS companions are found in the data, and unfortunately no IRDIS data were collected for this object. \\
    
    \item HD~\num{152200} is an O9.7~IV(n) star in NGC 6231 \citep{Sota2014}. It has been reported as an eclipsing binary with period of about 9 days by \citet{PosoNunez2019}. Close to the central star, at a separation of 0\farcs72 we found a  0.1 \Msun~star that is 5.2 Myrs old. The spectrum of the companion is shown in Fig.\ref{Fig: HD152200B spectrum}. \\

\begin{figure}
   \centering
   \includegraphics[width=\columnwidth]{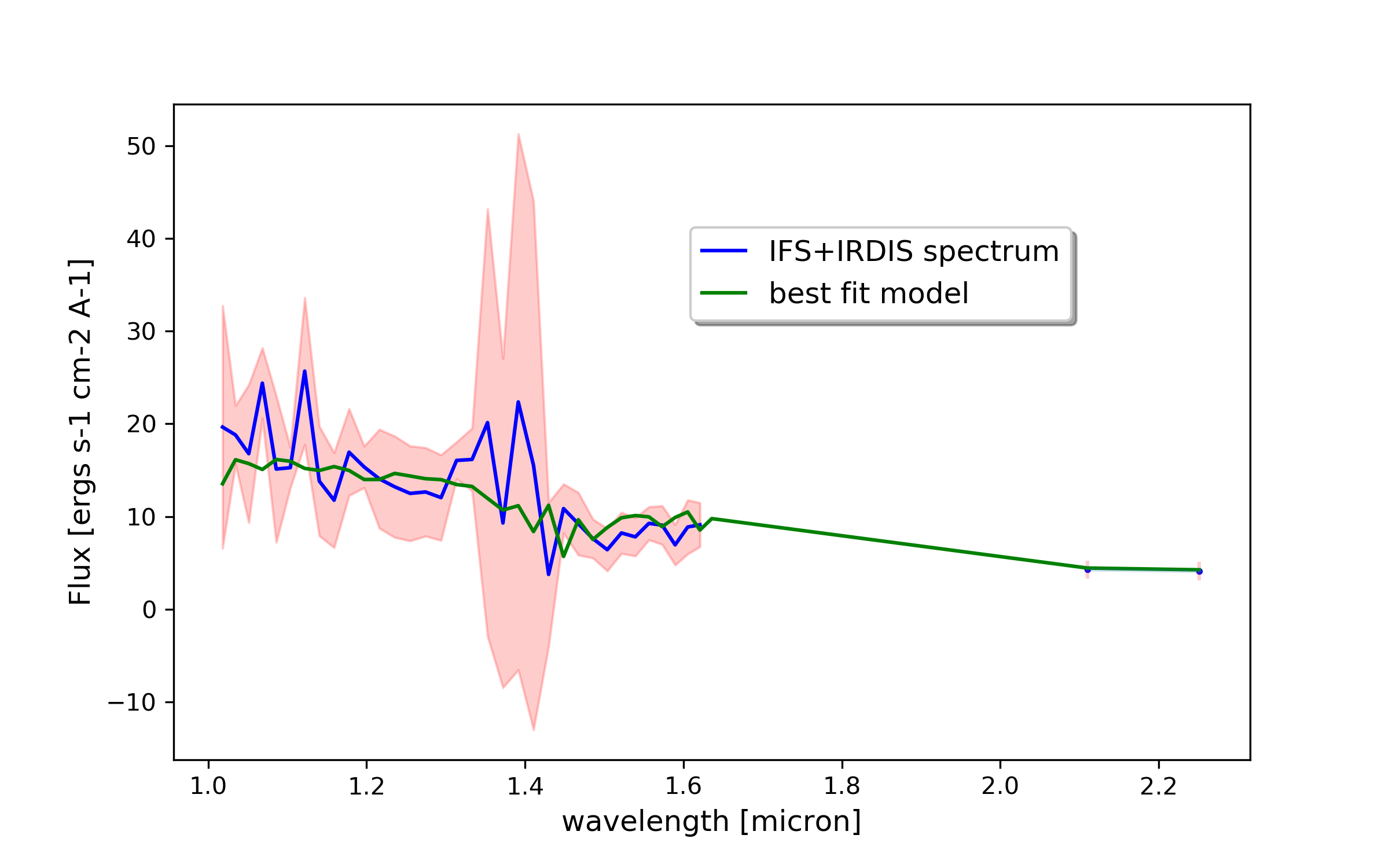}
      \caption{Observed spectrum for HD~\num{152200}~B. }
         \label{Fig: HD152200B spectrum}
\end{figure}
    
\end{itemize}

\subsection{Sensitivity limits}\label{sec:Sensitivity limits}
\begin{figure}
   \centering
   \includegraphics[width=\columnwidth]{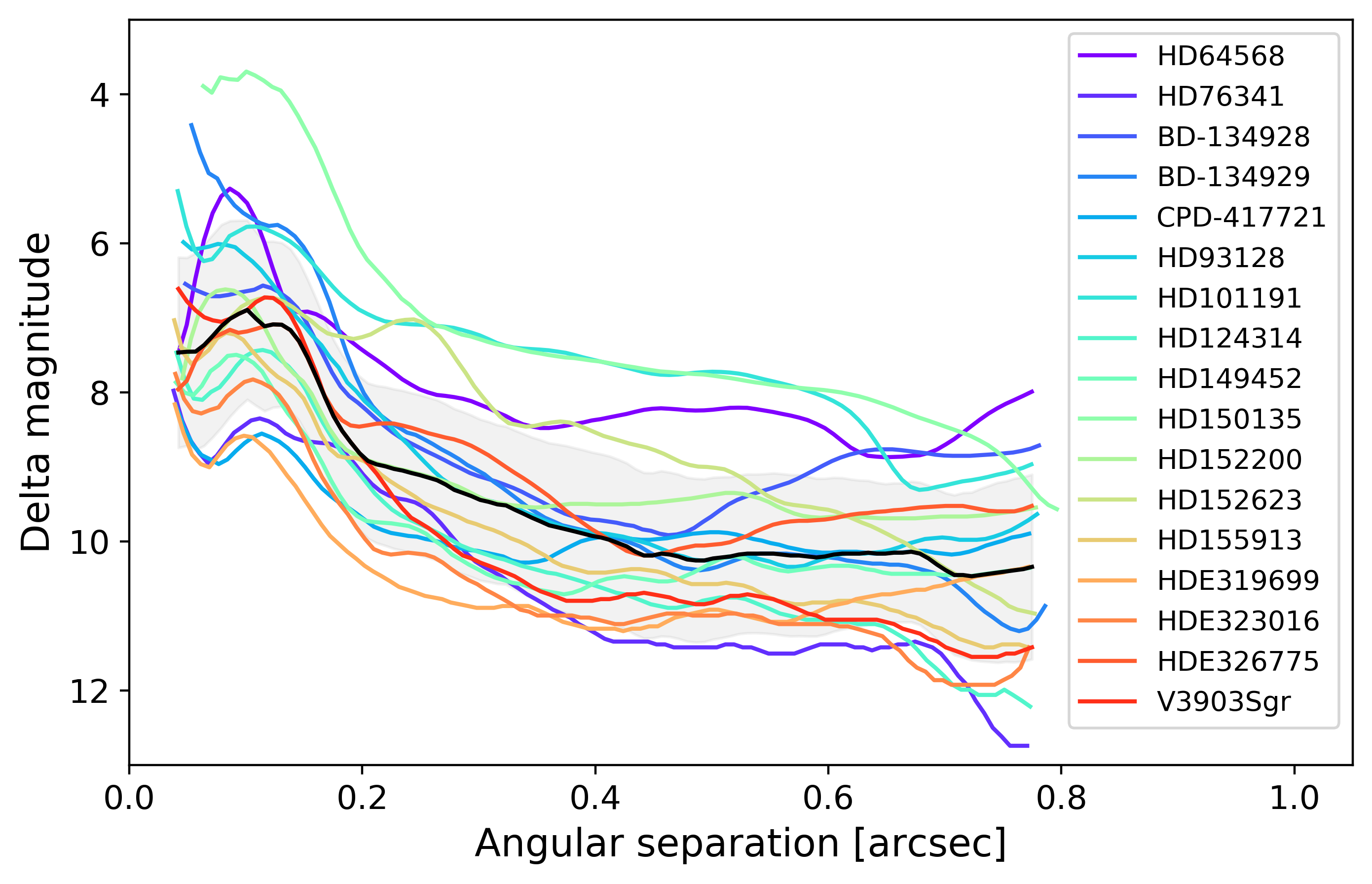}
      \caption{Sensitivity of our IFS observations expressed as magnitude difference vs. angular separation to the central star. The curves for each objects correspond to the 5-$\sigma$ contrast.}
         \label{Fig: IFS contrasts}
   \end{figure}

\begin{figure*}
   \centering
   \includegraphics[width=\columnwidth]{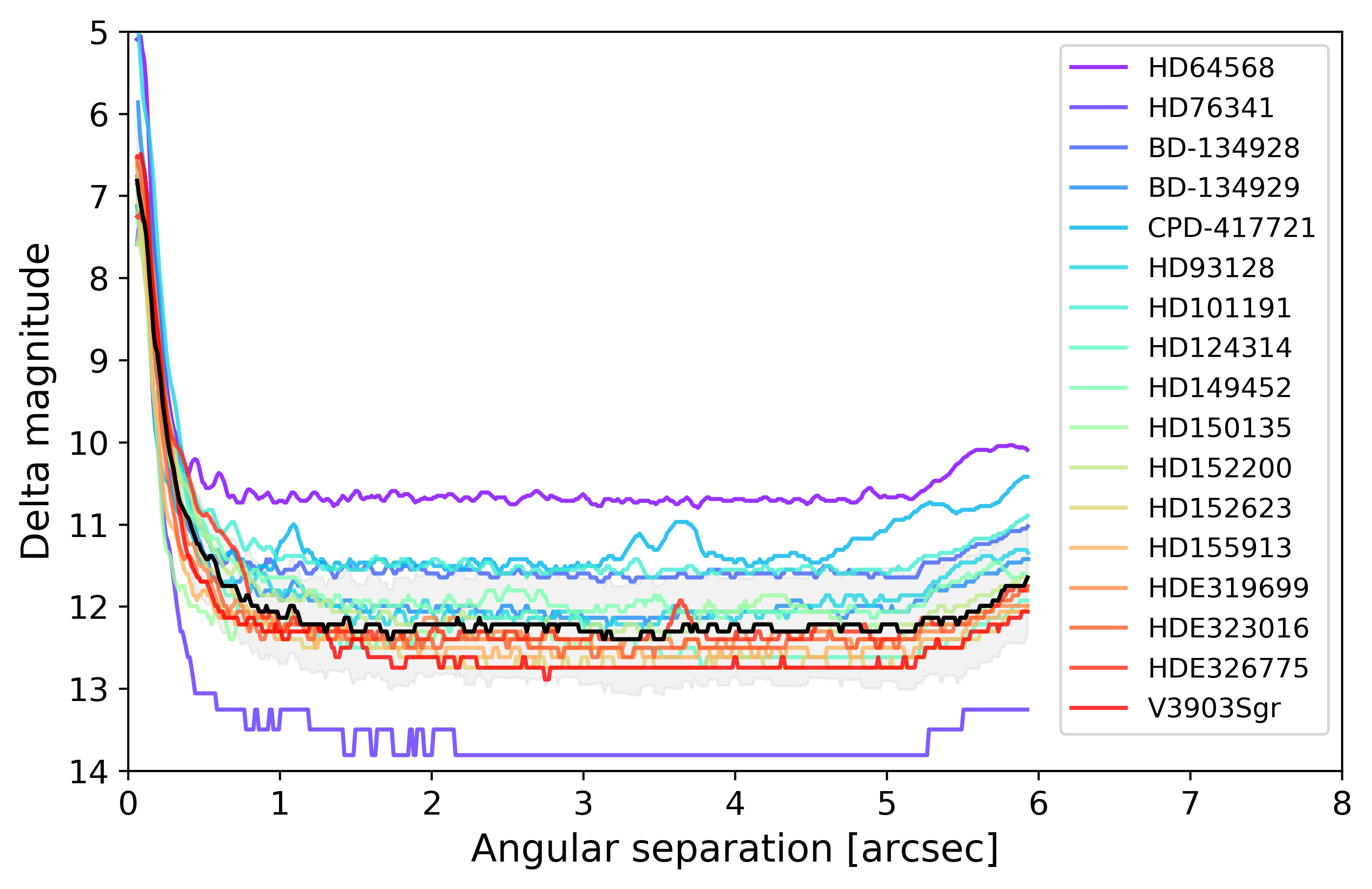}
   \includegraphics[width=\columnwidth]{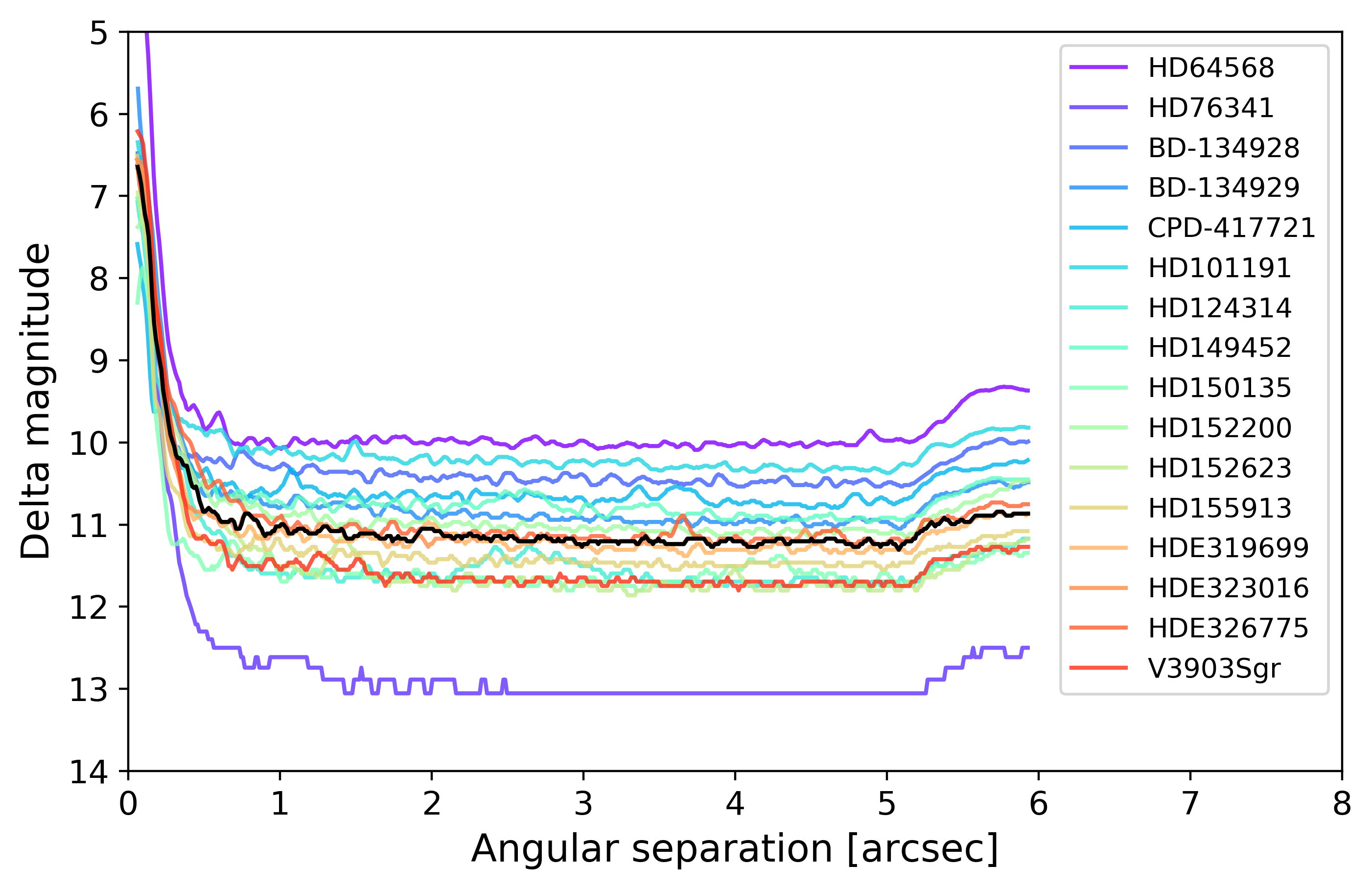}
      \caption{5-$\sigma$ contrast curves for our observations in K1 (left) and K2 (right) IRDIS bands.}
         \label{Fig: IRDIS contrasts}
   \end{figure*}

We estimated the sensitivity we reached with our observations in terms of magnitude difference as a function of the angular separation to the central star. We used the VIP contrast curve function that calculates the contrast limits for a chosen $\sigma$ level by injecting artificial companions (with scaled flux based on the unsaturated PSF of the central object) and calculated the noise and the algorithm throughput at different radial distances from the center. At close separations, we took into account the small sample statistics correction proposed in \citep{Mawet2014}. The 5-$\sigma$ sensitivity curves that we obtained for each target and for both IFS and IRDIS observations are presented in Figs.~\ref{Fig: IFS contrasts} and~\ref{Fig: IRDIS contrasts}, respectively.

\section{Discussion}\label{Sec:discussion}
\begin{figure}
   \centering
   \includegraphics[width=\columnwidth]{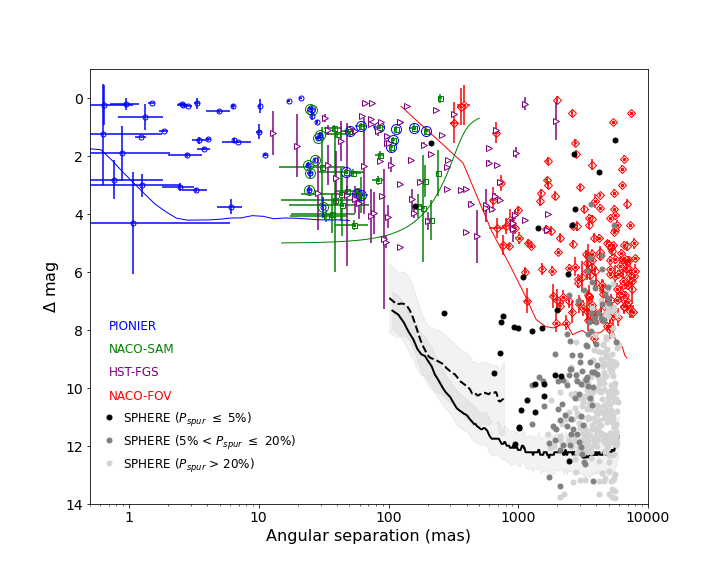}
      \caption{ $\Delta mag$ vs. separation diagram. The location of the newly discovered sources (black and grey dots), as well as the median IRDIS and IFS contrast limits are compared with the outcome of previous surveys \citep{Sana2014}. Black, grey, and light-grey dots corresponds to sources with $P_\mathrm{spur} \leq 5$\%, $5$\% $< P_\mathrm{spur} \leq 20$\%, and $P_\mathrm{spur}> 20$\%, respectively.} 
         \label{Fig: SPHERE detections}
   \end{figure}

All the sources discovered in this work are presented in the $\Delta mag$ vs. separation plot shown in Fig~\ref{Fig: SPHERE detections}. Despite the modest number of stars observed in this study compared to the SMASH+ survey, the IFS discoveries are clearly populating a  region of the parameter space that has never been reached before. 

Among them, four are newly detected sources and have estimated masses below 0.2 \Msun. Ranging from  mass ratios of q=0.002 (0.004) to 0.01, these estimates makes them the lowest mass-ratio companions ever discovered around O-type stars.
These objects are located between 400-1500 AU, based on the distance of the central stars.  
Such separations are in agreement with recent observations of fragments and substructures in Keplerian disks around (proto-)O stars \citep[e.g.,][]{Ilee2018,Beuther2017, Maud2019}. Whether these substructures and fragments will survive,  end up as companions at such large separations, or migrate inwards to form spectroscopic binary systems is still an open question that current hydrodynamics simulations are working to address \citep{Oliva2020}. 

Following the definitions given in \cite{Sana2014}, we can calculate the fraction of companions as the mean number of companions per central star, that is, the ratio of the total number of companions to the sample size. The error on the fraction can be estimated with Poisson statistics \citep[see equation 9 in][]{Sana2014}.
Assuming that all sources detected with angular separations of less than 0\farcs9 are physically bound companions, the observed (uncorrected for bias) fraction of companions for O-type stars between 150 and 900 mas (based on the effective size of the IFS fov) is $0.39\pm0.15$. 
If we take into account the spurious association probability for sources with $P_\mathrm{spur} \leq 5$\% in the larger IRDIS field of view (FOV), this fraction increases to $1.6\pm0.3$ in the separation range from 0\farcs9 to 6\arcsec. 
In order to compare our results with those of the SMASH+ survey, we need to restrict the comparison over the delta-magnitude and separation ranges covered by both studies. For angular separations between 0\farcs15 and 0\farcs9, and contrasts $<$4 mag, we observed a fraction of companions of 0.12$\pm$0.07, whereas in the angular separation range of 0\farcs9-6\arcsec~and $\Delta$mag$<$8, we obtained 0.76$\pm$0.21.
Due to a brightness limitation, the SMASH+ survey only observed a small fraction of O-type dwarfs. For this subset of objects (50 in total), over the same ranges, they observed a companion fraction of 0.24$\pm$0.07 between 0\farcs15 and 0\farcs9 and of 0.94$\pm$0.14 between 0\farcs9-6\arcsec. Both fractions are consistent with our findings within the errors. The correction for observational and selection biases goes beyond the scope of the present work and it will be presented in an upcoming paper.

If we consider as bound objects all sources with $P_\mathrm{spur} \leq 5$\%, the mean number of companions combining spectroscopic and eclipsing as well as visual multiples is 2.3. This number also includes known runaway stars. This implies not only that most massive stars are in multiple systems but also that triple or higher-order systems are more common than simple binaries. This outcome is in agreement with the results from the MONOS (Multiplicity Of Northern O-type Spectroscopic systems) project \citep{MaizApellaniz2019}.

Finally, concerning the influence of the environment density on the companion fraction, we do not see any strong correlation on the total number of companions in the 0\farcs15-6\arcsec~separation range, when comparing stars in denser cluster (e.g., Trumpler 14), OB associations, or rather isolated objects (see Table~\ref{table:Detection summary}). Nevertheless, we note that the previously reported runaway stars HD~\num{64568} and HD~\num{155913} are confirmed to be single, according to our spurious association probability criterion, as well.

 \begin{table*}[t]
 \caption[]{\label{table:Detection summary} Summary of detected companions ($P_\mathrm{spur} \leq 5$\%) in IFS and IRDIS for all objects. }
\begin{tabular}{lccccc}
 \hline \hline
  Object ID& \# known spectroscopic/eclipsing comp. & \# comp. IFS &  \# comp. IRDIS & total & Cluster/Association \\
 \hline
HD~\num{64568}  &  --  & --  & -- & 0 & NGC 2467, Pup OB1 \\
HD~\num{93128}  &  --  & -- & 2 & 2 & Trumpler 14, Car OB1 \\
HD~\num{155913}  &  --    & -- & -- & 0 & RCW 114 (possible runaway)\\
HDE~\num{319699}  & --     & -- & 6 & 6 & NGC 6334,RCW 127 \\
HD~\num{124314}  &   1   & -- & 4 & 5 &  --  \\
HD~\num{150135}   &  1   & -- & 3 & 4 & NGC 6193, Ara OB1a \\
HDE~\num{326775}  &  --   & -- & 4 & 4 & RCW 113 \\
V3903~Sgr  &   1  & -- & 1 & 2 & Sgr OB1  \\
HD~\num{152623}  &  1   & 1 & 2 & 4 & Trumpler 24, Sco OB1 \\
BD~$-$13\degr~4929   &  2   &  3 & 0 & 5 & NGC 6611 \\
HD~\num{101191}  &  --    & -- & -- & 0 & IC 2944, Cru OB1 \\
HDE~\num{323016}  &  --    & -- & -- & 0 &  --  \\
HD~\num{149452}  &   --   & -- & 1 & 1 & RCW 108, Ara OB1ab \\
HD~\num{76341}  &  --   & 1 & 2 & 3 & RCW, VelOB1   \\
BD~$-$13\degr~4928  &   --   & -- & 1 & 1 & NGC 6611 \\
CPD~$-$41\degr~7721  &   --   & 1 & 3 & 4 & NGC 6231, Sco OB1 \\
HD~\num{123056}  &  2   & -- & -- & 2 &  --  \\
HD~\num{152200}  &   1   & 1 & 1 & 3 & NGC 6231,  Sco OB1 \\
\hline
\end{tabular}
\end{table*}

\section{Conclusions}\label{Sec:conclusions}
In this work we used VLT/SPHERE in IRDIFS\_EXT mode to simultaneously carry out observations with the  IFS and IRDIS subsystems and characterize the multiplicity properties of a sample of 18 O-type stars from stellar clusters and loose associations between 0\farcs15 and  6\arcsec.
 We summarize the main results of our study below.

   \begin{enumerate}
      \item Despite the small size of the sample, compared to previous  high angular resolution surveys \citep[e.g.,  SMaSH+,  ][]{Sana2014}, we added a considerable number of companions in the 0\farcs15-1\farcs5 angular separation range. By reaching $\Delta$H=12, we also opened up a completely new region of the parameter space, with the possibility of exploring the existence of sub-solar companions around massive O-type stars.
      \item We found and characterized seven (five  of  which are previously unknown) companions  within 0\arcsec9 from  the  central star.  The five newly discovered companions have estimated masses below 0.25 \Msun, making them the highest mass-ratio binaries or multiple systems known thus far. 
      \item In addition to the close stellar companions, we detected several other sources in the larger IRDIS FoV with $P_\mathrm{spur}< 5$\%. Although we expect many of them  to  be physically bound companions, only future proper motion observations  will enable  us to confirm their companionship.\
      \item If we assume that all sources with angular separations below 0\farcs9 are physically bound companions, and by taking into account the spurious association probability for those with $P_\mathrm{spur} \leq 5$\% from 0\farcs9 to 6\arcsec, the observed (uncorrected for bias) fraction of companions for O-type stars is $0.39\pm0.15$  from 0\farcs15 to 0\farcs9 and $1.6\pm0.3$ in the separation range from 0\farcs9 to 6\arcsec. These fractions clearly support the idea that massive stars form almost exclusively in multiple systems, with preference for triples or higher-order multiples.
      \item The results of this study demonstrate that probing extreme contrasts as allowed by large AO-assisted coronagraphic surveys is fundamental to fully constrain the multiplicity properties of massive star companions in regions of the parameter space that remained unexplored so far and to characterize the low-mass end of the mass function.

   \end{enumerate}

\begin{acknowledgements}
  The authors thank the referee, Jes{\'u}s Ma{\'\i}z Apell{\'a}niz, for their detailed and careful comments that helped to improve this manuscript.
  This work is based on observations collected at the European Southern Observatory under programs ID 60.A-9369(A) and 095.D-0495(A). We thank the SPHERE Data Centre, jointly operated by OSUG/IPAG (Grenoble), PYTHEAS/LAM/CeSAM (Marseille), OCA/Lagrange (Nice) and Observatoire de Paris/LESIA (Paris) and supported by a grant from Labex OSUG@2020 (Investissements d'avenir a ANR10 LABX56). We especially thank P. Delorme, E. Lagadec and J. Milli (SPHERE Data Centre) for their help during the data reduction process.

  We acknowledge support from the FWO 1280121N grant and the FWO-Odysseus program under project G0F8H6N. This project has further received funding from the European Research Council under European Union's Horizon 2020 research program (grant agreement No 772225, MULTIPLES) and under the European Union's Seventh Framework Program (ERC Grant Agreement n. 337569).
\end{acknowledgements}


\bibliographystyle{aa} 
\bibliography{references} 

\end{document}